\documentstyle[12pt]{article}
\newlength{\extraspace}
\newlength{\extraspaces}
\newcommand{\be}{\begin{equation}
\addtolength{\abovedisplayskip}{\extraspaces}
\addtolength{\belowdisplayskip}{\extraspaces}
\addtolength{\abovedisplayshortskip}{\extraspace}
\addtolength{\belowdisplayshortskip}{\extraspace}}
\newcommand{\ee}{\end{equation}}

\newcommand{\ba}{\begin{eqnarray}
\addtolength{\abovedisplayskip}{\extraspaces}
\addtolength{\belowdisplayskip}{\extraspaces}
\addtolength{\abovedisplayshortskip}{\extraspace}
\addtolength{\belowdisplayshortskip}{\extraspace}}
\newcommand{\ea}{\end{eqnarray}}

\newcommand{\newsection}[1]{
\vspace{15mm}
\pagebreak[3]
\addtocounter{section}{1}
\setcounter{equation}{0}
\setcounter{subsection}{0}
\setcounter{footnote}{0}
\begin{flushleft}
{\large\bf \thesection. #1}
\end{flushleft}
\nopagebreak
\medskip
\nopagebreak}

\newcommand{\Tr}{{\rm Tr}}

\begin{document}

\addtolength{\baselineskip}{.8mm}

{\thispagestyle{empty}
\noindent \hspace{1cm}  \hfill IFUP--TH/2002--24 \hspace{1cm}\\
\mbox{}                 \hfill June 2002 \hspace{1cm}\\

\begin{center}
\vspace*{1.0cm}
{\large\bf Remarks on the $U(1)$ axial symmetry in QCD} \\
{\large\bf at zero and non--zero temperature} \\
\vspace*{1.0cm}
{\large Enrico Meggiolaro}\\
\vspace*{0.5cm}{\normalsize
{Dipartimento di Fisica, \\
Universit\`a di Pisa, \\
Via Buonarroti 2, \\
I--56127 Pisa, Italy.}}\\
\vspace*{2cm}{\large \bf Abstract}
\end{center}

\noindent
This paper is organized in two parts.
The first part (Sections 2--5) is dedicated to the theory at $T=0$ and
contains a pedagogical review of some fundamental aspects related with
the chiral symmetries of QCD, the $U(1)$ problem and its solution
proposed by 'tHooft, Witten and Veneziano. In the second part (Sections
6--14) we discuss the role of the $U(1)$ axial symmetry for the phase
structure of QCD at finite temperature. One expects that, above a certain
critical temperature, also the $U(1)$ axial symmetry will be restored.
We will try to see if this transition has (or has not) anything to do
with the usual chiral transition: various possible scenarios are discussed.
In particular, we analyse a scenario in which the $U(1)$ axial symmetry is
still broken above the chiral transition. We will show that this scenario
can be consistently reproduced in the full respect of the relevant QCD
Ward Identities and also using an effective Lagrangian model.
A new order parameter is introduced for the $U(1)$ axial symmetry.
}
\vfill\eject

\newsection{Introduction}

\noindent
It is generally believed that a phase transition which occurs in QCD at a
finite temperature is the restoration of the spontaneously broken 
$SU(L) \otimes SU(L)$ chiral symmetry in association with $L$ massless quarks.
At zero temperature the chiral symmetry is broken spontaneously by the
condensation of $q\bar{q}$ pairs and the $L^2-1$ $J^P=0^-$ mesons
are just the Nambu--Goldstone (NG) bosons associated with this breaking
\cite{current-algebra,quark-model,Nambu60,Chou61,Goldstone61}.
At high temperatures the thermal energy breaks up the $q\bar{q}$ condensate,
leading to the restoration of chiral symmetry. We expect that this property
not only holds for massless quarks but also continues for a small mass region.
The order parameter for the chiral symmetry breaking is apparently 
$\langle \bar{q}q \rangle \equiv \sum_{i=1}^L \langle \bar{q}_i q_i \rangle$: 
the chiral symmetry breaking corresponds to the non--vanishing of 
$\langle \bar{q}q \rangle$ in the chiral limit $\sup(m_i) \to 0$.
From lattice determinations of the chiral order parameter
$\langle \bar{q}q \rangle$ one knows that the $SU(L) \otimes SU(L)$ chiral
phase transition temperature $T_{ch}$, defined as the temperature at which
the chiral condensate $\langle \bar{q}q \rangle$ goes to zero (in the chiral
limit $\sup(m_i) \to 0$), is nearly equal to the deconfining temperature
$T_c$ (see, e.g., Ref. \cite{Blum-et-al.95}).
But this is not the whole story: QCD possesses not only an 
approximate $SU(L) \otimes SU(L)$ chiral symmetry, for $L$ light quark 
flavours, but also a $U(1)$ axial symmetry (at least at the classical 
level) \cite{Weinberg75,tHooft76}.
The role of the $U(1)$ symmetry for the finite temperature phase 
structure has been so far not well studied and it is still an open question
of hadronic physics whether the fate of the $U(1)$ chiral symmetry of QCD has
or has not something to do with the fate of the $SU(L) \otimes SU(L)$ chiral
symmetry. In the following Sections we will try to answer these questions:
\begin{itemize}
\item{} At which temperature is the $U(1)$ axial symmetry restored?
(if such a critical temperature does exist!)
\item{} Does this temperature coincide with the deconfinement temperature 
and with the temperature at which the $SU(L)$ chiral symmetry is restored?
\end{itemize}
In the ``Witten--Veneziano mechanism'' \cite{Witten79a,Veneziano79}
for the resolution of the $U(1)$ problem, a fundamental role is played by
the so--called ``topological susceptibility'' in a QCD without
quarks, i.e., in a pure Yang--Mills (YM) theory, in the large--$N_c$ limit
($N_c$ being the number of colours):
\be
A = \displaystyle\lim_{k \to 0}
\displaystyle\lim_{N_c \to \infty}
\left\{ -i \displaystyle\int d^4 x e^{ikx} \langle T Q(x) Q(0) \rangle
\right\} ,
\label{eqn1}
\ee
where $Q(x) = {g^2 \over 64\pi^2}\varepsilon^{\mu\nu\rho\sigma} F^a_{\mu\nu}
F^a_{\rho\sigma}$ is the so--called ``topological charge density''.
This quantity enters into the expression for the squared mass of the $\eta'$:
$m^2_{\eta'} = {2L A \over F^2_\pi}$, where $L$ is the number of light quark
flavours taken into account in the chiral limit.
Therefore, in order to study the role of the $U(1)$ axial symmetry for the
full theory at non--zero temperatures, one should consider the YM topological
susceptibility $A(T)$ at a given temperature $T$, formally defined as
a large--$N_c$ limit of a certain expectation value $\langle \ldots \rangle_T$
in the full theory at the temperature $T$ \cite{EM1998}:
\be
A(T) = \displaystyle\lim_{k \to 0}
\displaystyle\lim_{N_c \to \infty}
\left\{ -i \displaystyle\int d^4 x e^{ikx} \langle T Q(x) Q(0) \rangle_T
\right\} .
\label{EQN1.3}
\ee
In other words, $A(T) = \displaystyle\lim_{k \to 0} A_0(k,T) = A_0(0,T)$,
where the quantity
\be
A_0(k,T) =
\displaystyle\lim_{N_c \to \infty}
\left\{ -i \displaystyle\int d^4 x e^{ikx} \langle T Q(x) Q(0) \rangle_T
\right\} ,
\label{EQN1.4}
\ee
at four--momentum $k$ and physical temperature $T$, is nothing but the
leading--order term in the $1/N_c$ expansion of the corresponding quantity
\be
\chi(k,T) =
-i \displaystyle\int d^4 x e^{ikx} \langle T Q(x) Q(0) \rangle_T ,
\label{EQN1.5}
\ee
of the full theory, at the same four--momentum $k$ and physical temperature
$T$. [From the general theory of $1/N_c$ expansion it is known that
$A_0(k,T)$ is of order ${\cal O}(1)$, while successive terms coming
from including fermions, are of order ${\cal O}(1/N_c)$.]
It is in this sense that one can talk about the behaviour of $A(T)$
above or below the chiral transition temperature $T_{ch}$.

The problem of studying the behaviour of $A(T)$ as a function of the
temperature $T$ was first addressed, in lattice QCD,
in Refs. \cite{Teper86,EM1992a,EM1995b}.
Recent lattice results \cite{Alles-et-al.97} (obtained for the $SU(3)$
pure--gauge theory) show that the YM topological susceptibility $A(T)$
is approximately constant up to the critical temperature $T_c \simeq T_{ch}$,
it has a sharp decrease above the transition, but it remains different
from zero up to $\sim 1.2~T_c$.
In the Witten--Veneziano mechanism \cite{Witten79a,Veneziano79},
a (no matter how small!) value different from zero for $A$ is related to the
breaking of the $U(1)$ axial symmetry, since it implies the existence of a 
pseudo--Goldstone particle with the same quantum numbers of the $\eta'$
(see also Ref. \cite{Veneziano80}).
Therefore, the available lattice results for the topological susceptibility
show that the $U(1)$ chiral symmetry is restored at a temperature $T_{U(1)}$
greater than $T_{ch}$.

Another way to address the same question is to look at the behaviour at
non--zero temperatures of the susceptibilities related to the
propagators for the following meson channels \cite{Shuryak94}
(we consider for simplicity the case of $L=2$ light flavours):
the isoscalar $I=0$ scalar channel $\sigma$ (also known as $f_0$ in the
modern language of hadron spectroscopy), interpolated by the operator
$O_\sigma = \bar{q} q$;
the isovector $I=1$ scalar channel $\delta$
(also known as $a_0$), interpolated by the operator
$\vec{O}_\delta = \bar{q} {\vec{\tau} \over 2} q$;
the isovector $I=1$ pseudoscalar channel $\pi$, interpolated by the operator
$\vec{O}_\pi = i\bar{q} \gamma_5 {\vec{\tau} \over 2} q$;
the isoscalar $I=0$ pseudoscalar channel $\eta'$, interpolated by the operator
$O_{\eta'} = i\bar{q} \gamma_5 q$.
Under $SU(2)_A$ transformations, $\sigma$ is mixed with $\pi$: thus the
restoration of this symmetry at $T_{ch}$ requires identical correlators
for these two channels. Another $SU(2)$ chiral multiplet is $(\delta,\eta')$.
On the contrary, under the $U(1)_A$ transformations, $\pi$ is mixed
with $\delta$: so, a ``practical restoration'' of the $U(1)$ axial
symmetry should imply that these two channels become degenerate, with
identical correlators. Another $U(1)$ chiral multiplet is $(\sigma,\eta')$.
(Clearly, if both chiral symmetries are restored, then all $\pi$, $\eta'$,
$\sigma$ and $\delta$ correlators should become the same.)
In practice, one can construct, for each meson channel $f$, the
corresponding chiral susceptibility
\be
\chi_f = \displaystyle\int d^4x \langle O_f(x) O_f^\dagger(0) \rangle ,
\label{eqn2}
\ee
and then define two order parameters:
$\chi_{SU(2) \otimes SU(2)} \equiv \chi_\sigma - \chi_\pi$, and
$\chi_{U(1)} \equiv \chi_\delta - \chi_\pi$.
If an order parameter is non--zero in the chiral limit, then the
corresponding symmetry is broken.
Present lattice data for these quantities seem to indicate that the $U(1)$
axial symmetry is still broken above $T_{ch}$, up to $\sim 1.2~T_{ch}$,
where the $\delta$--$\pi$ splitting is small but still
different from zero \cite{Bernard-et-al.97,Karsch00,Vranas00}.
In terms of the left--handed and right--handed quark fields
[$q_{L,R} \equiv {1 \over 2} (1 \pm \gamma_5) q$], one has the following
expression for the difference between the correlators for
the $\delta^+$ and $\pi^+$ channels:
\ba
\lefteqn{
{\cal D}_{U(1)}(x) \equiv \langle O_{\delta^+}(x) O_{\delta^+}^\dagger(0)
\rangle - \langle O_{\pi^+}(x) O_{\pi^+}^\dagger(0) \rangle } \nonumber \\
& & = 2 \left[ \langle \bar{u}_R d_L(x) \cdot \bar{d}_R u_L(0) \rangle
+ \langle \bar{u}_L d_R(x) \cdot \bar{d}_L u_R(0) \rangle \right] .
\label{eqn3}
\ea
(The integral of this quantity, $\int d^4x {\cal D}_{U(1)}(x)$, is just equal
to the $U(1)$ chiral parameter $\chi_{U(1)} = \chi_\delta - \chi_\pi$.)
What happens below and above $T_{ch}$?
Below $T_{ch}$, in the chiral limit $\sup(m_i) \to 0$, the left--handed
and right--handed components of a given light quark flavour ({\it up} or
{\it down}, in our case with $L=2$) can be connected through the
quark condensate, giving rise to a non--zero contribution to the
quantity ${\cal D}_{U(1)}(x)$ in Eq. (\ref{eqn3}) (i.e., to the quantity
$\chi_{U(1)}$).
But above $T_{ch}$ the quark condensate is zero: so, how can the quantity
${\cal D}_{U(1)}(x)$ (i.e., the quantity $\chi_{U(1)}$) be different from zero
also above $T_{ch}$, as indicated by present lattice data?
The only possibility in order to solve this puzzle seems to be that of
requiring the existence of a genuine four--fermion local condensate,
which is an order parameter for the $U(1)$ axial symmetry and which
remains different from zero also above $T_{ch}$.
This will be discussed in Section 7.

The paper is organized in two parts.
The first part (Sections 2--5) is dedicated to the theory at $T=0$ and
contains a pedagogical review of some fundamental aspects related with
the chiral symmetries of QCD (Section 2), the $U(1)$ problem and its solution
proposed by 'tHooft (Section 3), the Witten's mechanism (Section 4) and the
Veneziano's mechanism (Section 5). 
In the second part (Sections 6--14) we discuss the role of the $U(1)$ axial
symmetry for the phase structure of QCD at finite temperature.
One expects that, above a certain critical temperature $T_{U(1)}$, also the
$U(1)$ axial symmetry will be restored. We will try to see if this transition
has (or has not) anything to do with the usual chiral transition: various
possible scenarios are discussed in Section 6.
In particular, we analyse a scenario in which the $U(1)$ axial symmetry is
still broken above the chiral transition \cite{EM1994a,EM1994b,EM1994c}.
In Section 7 a new order parameter is introduced for the $U(1)$ axial symmetry.
In Section 8 we make an analysis of the relevant QCD Ward Identities (WI)
in the range of temperatures between $T_{ch}$ and $T_{U(1)}$.
The saturation of the relevant QCD Ward Identities is obtained in a
$1/N_c$ expansion. A new $2L$--fermion effective vertex is introduced to
take care of the new order parameter for the $U(1)$ chiral symmetry alone.
In Sections 8 and 9 we introduce a new meson field associated with the
new $U(1)$ chiral order parameter and then, in Section 9, we study the
form of a new effective Lagrangian including, in addition to the usual
quark--anti-quark meson fields $U_{ij}$, also the new $2L$--fermion field $X$.
In Sections 10 and 11 we study the mass spectrum of the theory respectively 
above and below the $SU(L) \otimes SU(L)$ chiral phase transition.
In Section 12 we apply the results of Section 11 to the ``{\it real--world}''
case (at $T < T_{ch}$), where there are $L=3$ light flavours, named $u$, $d$
and $s$: we derive a {\it generalized Witten--Veneziano formula} for the
$\eta'$ mass. In Section 13 we show that the above results are
``model--independent'', i.e., independent on additional terms in the Lagrangian.
In Section 14 we analyse the consequences of our theoretical model on the 
chiral condensate $\langle \bar{q} q \rangle$ and on the topological
susceptibility $\chi$, in the {\it full} theory with quarks.
In particular we are interested in their dependence on the light quark masses
(something about this is also said in the Appendix).
In Section 14 we also discuss the $N_c$--dependence for some relevant
quantities that we have found.
Finally, the conclusions and an outlook are given in Section 15.

\newsection{The chiral symmetries of QCD}

\noindent
The parameter $\Lambda_{QCD} \sim 0.5$ GeV (in the $\overline{MS}$ 
renormalization scheme), associated with asymptotic 
freedom, defines a high--momentum (or short--distance) regime
($k^2/\Lambda^2_{QCD} \gg 1$) in which quarks and gluons can be treated as 
weakly interacting particles in perturbative QCD. At the other end of the 
scale ($k^2/\Lambda^2_{QCD} \le 1$) is low--energy hadronic physics, in 
which the interacting units are not individual quarks and gluons, but 
hadrons. So far we cannot solve QCD in this domain. However, from the vast 
amount of experimental information available, some highly fruitful ideas 
had been developed long before QCD was invented. Translated into expected 
properties of QCD, some of these ideas become concrete statements that are 
much easier to comprehend.

Perhaps the most relevant example of this is represented by the chiral 
symmetries of strong interactions. They were known since the $60$'s, when 
the only theoretical instruments available for their analysis were
{\it current algebra} \cite{current-algebra} and a rudimental 
{\it quark model} \cite{quark-model}.
Translated into the QCD language, they become a much more clear and
well-defined subject. Let us consider, in fact, the QCD Lagrangian:
\be
{\cal L}_{QCD} = -{1 \over 4}F^a_{\mu\nu}F^{a\mu\nu} +
\bar{q}i\gamma^\mu D_\mu q -\bar{q} M q ,
\label{EQNA1}
\ee
where $M$ is the quark mass matrix:
\be
M = \pmatrix{ m_u& & & \cr
& m_d& & \cr
& & m_s& \cr
& & & \ddots \cr}
.
\label{EQNA2}
\ee
Since the quark masses are different from one another, the symmetry group 
of this theory is $G = U(1)_u \otimes U(1)_d \otimes \ldots$ There is one 
conserved charge for every one of the quark flavours, the corresponding 
conserved currents being $\bar{u}\gamma^\mu u$, $\bar{d}\gamma^\mu d$,
etc. If we take $L$ of these flavours to have zero masses (the physical 
meaning of this will be explained below \cite{Nambu60,Chou61}) 
the symmetry group of the theory 
becomes larger. The Lagrangian is invariant under independent rotations 
of the left-- and right--handed quark fields:
\be
q_L \to V_L q_L ~~~ , ~~~ q_R \to V_R q_R ,
\label{EQNA3}
\ee
where $q$ stands for the vector $(q_1,q_2,\ldots ,q_L)$, while
$q_L \equiv {1 \over 2}(1 + \gamma_5)q$ and
$q_R \equiv {1 \over 2}(1 - \gamma_5)q$,
with $\gamma_5 \equiv -i \gamma^0 \gamma^1 \gamma^2 \gamma^3$,
denote respectively the {\it left--handed} and the {\it right--handed} quark
fields; $V_L$ and $V_R$ are arbitrary $L \times L$ unitary matrices.
So the symmetry group is:
\be
G = SU(L)_L \otimes SU(L)_R \otimes U(1)_L \otimes U(1)_R .
\label{EQNA4}
\ee
We do not consider any more the part of the Lagrangian depending on the
``{\it heavy}'' quark fields.

There are two ways in which a continuous global symmetry of the Lagrangian 
may be realized in the vacuum:
\begin{itemize}
\item{} The {\it Wigner--Weyl} realization, in which all the generators 
of the symmetry group annihilate the vacuum. Then the physical states can 
be classified according to the irreducible unitary representations of the 
group.
\item{} The {\it Nambu--Goldstone} realization, in which some generators 
do not annihilate the vacuum. Then Goldstone theorem \cite{Goldstone61} 
states that, for each generator which does not annihilate the vacuum, there 
exists a spin--zero massless particle (Goldstone boson), with the quantum 
numbers of this generator and transforming according to the irreducible 
representations of the unbroken subgroup.
\end{itemize}
In nature $U(1)_{L+R}$ is realized in the Wigner--Weyl way and manifests 
itself directly as baryon--number conservation. On the contrary
$SU(L)_L \otimes SU(L)_R$ cannot be realized in such a way, because this 
would imply that each hadron multiplet is accompanied by a mirror multiplet 
of the same mass, but with opposite parity. No hint of this can be found in 
the hadronic spectrum. For example, there is not even an approximate mirror 
image of the nucleon iso--doublet. Thus one generally assumes that this 
symmetry is spontaneously broken down to:
\be
H = SU(L)_{L+R} ,
\label{EQNA5}
\ee
the vacuum being invariant only under the subgroup generated by the vector 
currents $\bar{q}\gamma^\mu {1 \over 2}\lambda_a q$. As a consequence, the 
spectrum of QCD (with $L$ massless quarks) must contain $L^2-1$ Goldstone 
bosons. Their quantum numbers can be read off from those of the states
$Q^A_a |0\rangle$, which result if the axial charge operators:
\be
Q^A_a = \int d^3{\bf x} A^0_a (x) ~~~ ; ~~~
A^\mu_a = \bar{q}\gamma^\mu \gamma_5 {1 \over 2}\lambda_a q ,
\label{EQNA6}
\ee
are applied to the vacuum: $J^P = 0^{-}$.
In reality, of course, the symmetry is explicitly broken by the quark--mass 
term in the Lagrangian, so that we do not observe in the hadronic spectrum 
the presence of massless Goldstone bosons.
Nevertheless, the eight lightest hadrons ($\pi^\pm ,\pi^0 ,K^\pm ,K^0 ,
\bar{K}^0 ,\eta$) are pseudoscalars and three of them ($\pi^\pm ,\pi^0$) 
are particularly light.
This pattern is to be expected if $m_u,~m_d,~m_s$ happen to be small 
when compared to the typical mass--scale of strong interactions,
i.e, $\Lambda_{QCD} \sim 0.5$ GeV (in the $\overline{MS}$ renormalization 
scheme). In such a way the Lagrangian is approximately invariant under
$SU(3)_L \otimes SU(3)_R$ and the spontaneous breakdown of this approximate 
symmetry generates $3^2-1=8$ approximate massless particles.
Moreover $m_u$ and $m_d$ must be small compared to $m_s$, so that the group
$SU(2)_L \otimes SU(2)_R$ is a much better symmetry of the Lagrangian. 
This is the reason why $2^2-1=3$ of the pseudoscalar mesons are 
particularly light \cite{Nambu60,Chou61}. 

In other words, the piece of the Lagrangian containing the lightest quark 
masses:
\be
\delta {\cal L}^{(mass)}_{QCD} =
-(m_u \bar{u}u + m_d \bar{d}d + m_s \bar{s}s ) ,
\label{EQNA7}
\ee
acts as a perturbative term ({\it ``chiral perturbation theory''}) with
respect to the remnant piece ${\cal L}^{(0)}_{QCD}$ of the Lagrangian:
${\cal L}^{(0)}_{QCD}$ is invariant under $SU(3)_L \otimes SU(3)_R$.
On the contrary, the mass parameters for the other quarks $c,~b,~t$ should 
all be much larger than $m_u,~m_d,~m_s$ ($m_c,~m_b,~m_t \ge \Lambda_{QCD}$),
because we see no trace of flavour $SU(4)$ or higher symmetries in the 
hadronic spectrum.

What is the dynamical reason for which $SU(3)_L \otimes SU(3)_R$ is broken 
in the Goldstone mode? We do not have an answer. In the 
non--renormalizable pre--QCD model of Nambu and Jona--Lasinio
\cite{Nambu-Jona-Lasinio61}, the cause of spontaneous symmetry breakdown is a 
direct nucleon--nucleon attraction built into the model, in analogy with 
the effective electron--electron attraction responsible for the formation 
of Cooper pairs in the theory of superconductivity. (In this theory the 
Goldstone boson is ``eated'' by the electromagnetic field through the Higgs 
mechanism and the photon becomes massive in a superconductor: this is 
called the {\it ``Meissner effect''}.)
A direct quark--quark interaction is ruled out by renormalizability in QCD; 
but an effective interaction could arise just as the effective 
electron--electron attraction in superconductivity arises from the more 
fundamental electron--phonon interaction. Maybe instantons play such a role 
in an effective quark--quark interaction \cite{Carlitz78}.

So far we have discussed what happens in nature of the symmetry
$SU(3)_L \otimes SU(3)_R \otimes U(1)_{L+R}$ of the QCD Lagrangian. However,
the symmetry group (\ref{EQNA4}) also contains the group $U(1)_{L-R}$, which
is usually called the {\it ``$U(1)$ chiral group''} or $U(1)_A$
($A$ stands for axial; the $U(1)_A$ group transformation is:
$q \to e^{-i\alpha \gamma_5}q$, $\bar{q} \to \bar{q}e^{-i\alpha\gamma_5}$).
The question of how $U(1)_A$ is manifested presents a puzzle, which is 
known as the {\it ``$U(1)$ problem''}. 

If the symmetry $U(1)_A$ were manifested directly in the Wigner--Weyl mode,
then in the chiral limit all massless hadrons would have a massless partner 
of opposite parity: but this is not the case in nature.
On the other hand, if it were broken in the Goldstone mode, then there 
should be an $I=0$ pseudoscalar Goldstone boson, whose perturbed state 
should have about the same mass as the pion. Using chiral perturbation 
theory, Weinberg \cite{Weinberg75} estimated its mass to be less than
$\sqrt{3} m_\pi$. Among the known hadrons, the only candidates with the 
right quantum numbers are $\eta (549)$ and $\eta' (958)$. Both violate the 
Weinberg bound: besides $\eta (549)$ has already been claimed by
the pion octet.

The $U(1)$ puzzle is: where is the extra Goldstone boson?

\newsection{The solution of the $U(1)$ problem}

\noindent
'tHooft \cite{tHooft76} removed the puzzle by observing 
that the $U(1)_A$ is not a symmetry of the theory at the quantum level, for 
the presence of instanton effects. We will not go into the detailed 
analysis but merely mention some relevant points.

The $U(1)$ axial current $J^{(L)}_{5, \mu} = \sum_{i=1}^L
\bar{q}_i \gamma_\mu \gamma_5 q_i$
(with $\gamma_5 \equiv -i \gamma^0 \gamma^1 \gamma^2 \gamma^3$)
is not conserved, due to a QCD axial anomaly (the so--called
{\it ``Adler--Bell--Jackiw (ABJ) anomaly''} \cite{Adler-Bell-Jackiw69}):
\be
\partial^\mu J^{(L)}_{5, \mu} = 2L Q(x) ,
\label{EQNB1}
\ee
where $Q(x) = {g^2 \over 64\pi^2}\varepsilon^{\mu\nu\rho\sigma}
F^a_{\mu\nu}F^a_{\rho\sigma}$ (with $\varepsilon^{0123} = 1$) is the
so--called {\it ``topological charge density''}, and $L$ is the number of
light quark flavours taken into account in the chiral limit ($L=2$ or $L=3$).
It is known that $Q(x)$ is a total $4$--divergence:
\be
Q(x) = \partial^\mu K_\mu (x) ~~ ; ~~
K_\mu = {g^2 \over 16\pi^2}\varepsilon_{\mu\alpha\beta\gamma}
A^\alpha_a \left( \partial^\beta A^\gamma_a + {1 \over 3}gf_{abc}
A^\beta_b A^\gamma_c \right) .
\label{EQNB2}
\ee
($f_{abc}$ are the group constants of $SU(3)_{colour}$).
The integral of $Q(x)$ over all space--time is called the {\it ``topological 
charge''}. We can define a conserved but non--gauge-invariant current:
\be
\tilde{J}^{(L)}_{5, \mu} = J^{(L)}_{5, \mu} - 2L K_\mu .
\label{EQNB3}
\ee
The generator of the $U(1)_A$ symmetry may be taken to be:
\be
\tilde{Q}^{(L)}_5 \equiv \int d^3{\bf x}\tilde{J}^{(L)}_{5, 0} =
\int d^3{\bf x} \left[ \displaystyle\sum_{i=1}^L q^\dagger_i \gamma_5 q_i
-2L K_0 \right] .
\label{EQNB4}
\ee
In the Abelian case $\tilde{Q}^{(L)}_5$ is gauge invariant, because of the
absence of topological charge. This is no longer true for the non--Abelian
case, and $\tilde{Q}^{(L)}_5$ is not a physical quantity. 
Moreover, $Q^{(L)}_5 \equiv \int d^3{\bf x} J^{(L)}_{5, 0}$ is not conserved,
because of the existence of instantons \cite{Belavin-et-al.75}. To see this, one
can integrate Eq. (\ref{EQNB1}) over the Euclidean four--space.
The result can be presented in the form:
\be
\displaystyle\int_{-\infty}^{+\infty} dt {d Q^{(L)}_5 \over dt} =
2L q[F] ,
\label{EQNB5}
\ee
where:
\be
q[F] = {g^2 \over 64\pi^2}\int d^4x \varepsilon_{\mu\nu\rho\sigma}
F^a_{\mu\nu}F^a_{\rho\sigma}
\label{EQNB6}
\ee
is the topological charge, a functional of the gauge field $F^a_{\mu\nu}$.
For $F^a_{\mu\nu}$ corresponding to one instanton, $q[F]=1$ (see
Ref. \cite{tHooft76}). Therefore, in this case, the 
boundary values of $Q^{(L)}_5$ in Euclidean time differ by:
\be
\Delta Q^{(L)}_5 = 2L q[F] .
\label{EQNB7}
\ee
This can be attributed to the fact that an instanton interpolates (in 
Euclidean time) between two gauge--field configurations differing by one 
unit of topological charge. Thus there seems to be no reason to expect 
$U(1)_A$ to be a symmetry.

However, the work of 'tHooft did remove the puzzle in its original form, 
but generated new questions. In particular: what becomes of the {\it 
would--be} Goldstone boson? Or, in other terms: why is the mass of the
$\eta' (958)$ (the {\it would--be} Goldstone boson of the $U(1)_A$) so large 
compared with the masses of the octet mesons?
Witten \cite{Witten79a} and Veneziano \cite{Veneziano79} proposed in 1979 a 
mechanism to answer these two questions and to give a reliable estimate of 
the $\eta'$ mass. This is now universally known as the 
{\it ``Witten--Veneziano mechanism''}.

\newsection{Witten's mechanism}

\noindent
Let us briefly review which was the idea of E. Witten \cite{Witten79a} to 
explain the large mass of the $\eta'$.
Starting from the resolution of the $U(1)$ problem given by 'tHooft in
1976 \cite{tHooft76} (which we have discussed in Section 3),
one may try to explain the large mass of the $\eta'$ by 
saying that the Goldstone boson receives a mass as a result of the anomaly. 
This statement has not a clear meaning, however, because there is 
apparently no way to turn off the anomaly, and therefore no sense in 
talking of the Goldstone boson that would have existed if there had been no 
anomaly. Witten observed, however, that in a certain sense it is possible 
to vary a parameter in QCD so as to turn off the anomaly: this parameter is 
the number $N_c$ of colours of the theory.

The $U(1)$ problem is then reconsidered from the point of view of the 
$1/N_c$ expansion (for a detailed analysis of this expansion see Refs.
\cite{tHooft74} and \cite{Veneziano76}). It is for large $N_c$ that the 
anomaly turns off and it is in the sense of the $1/N_c$ expansion that one 
can say that the anomaly gives a mass to a $U(1)$ particle that otherwise 
would have been massless.

There is also a more practical way to discuss the $U(1)$ problem.
Isgur \cite{Isgur76} and De R\`ujula, Georgi and 
Glashow \cite{DeRujula-et-al.75} 
suggested that in the $\eta'$, which is an $SU(3)$ singlet, the quark and 
antiquark can annihilate into gluons (this annihilation channel is absent 
for the non--singlet mesons). These quark--antiquark annihilation diagrams 
split the $\eta'$ from the light pseudoscalars. Of course this is not true 
at a given finite order of perturbation theory, because in finite orders of 
perturbation theory we have neither an $\eta'$ nor a $\pi$.
Yet in the context of the $1/N_c$ expansion, one can easily make the 
summation of an infinite class of Feynman diagrams, so as to generate bound 
states \cite{Witten79b}.
In the large--$N_c$ limit the $\eta'$ and $\pi$ are degenerate; this 
degeneracy is lifted by quark--antiquark annihilation diagrams that are 
suppressed by one power of $1/N_c$.

To see how the $\eta'$ gets a mass from the anomaly, we have to consider 
the Fourier transformed of the correlation function of the 
topological charge $Q(x)$:
\be
\chi(k) = -i\int d^4x e^{ikx} \langle TQ(x)Q(0) \rangle .
\label{EQNC1}
\ee
The $\chi(k)$ at zero momentum is called the {\it topological susceptibility}, 
which in the literature is usually indicated with $A$ or $\chi$:
\be
\chi = -i\int d^4x \langle TQ(x)Q(0) \rangle .
\label{EQNC2}
\ee
It is well known, from a direct analysis of the QCD Ward Identities
\cite{Crewther77}, that $\chi$ is zero in the theory with massless quarks.
This can also be seen if we write $\chi$ in the form (here $VT = \int d^4 x$
is an infinite four--volume which must be factorized):
\be
\chi = {1 \over VT} {i \over Z_\theta [\eta, \bar{\eta}]}
{d^2 Z_\theta [\eta ,\bar{\eta}] \over d\theta^2}
\displaystyle\vert_{\theta = 0 \atop \eta = \bar{\eta} = 0} ,
\label{EQNC3}
\ee
where $Z_\theta [\eta, \bar{\eta}]$ is the partition function of the theory 
including the so--called $\theta$--term (which in quarkless QCD labels 
different $\theta$--worlds that are physically distinct):
\ba
\lefteqn{
Z_\theta [\eta ,\bar{\eta}] =
\int [dA_\mu] [dq] [d\bar{q}] } \nonumber \\
& & \times  \exp \left\{ i\int d^4x \left[ 
-{1 \over 4}F^a_{\mu\nu}F^{a\mu\nu} + \bar{q} (i\gamma^\mu D_\mu - M) q
+ \bar{\eta} q + \bar{q} \eta + \theta Q(x) \right] \right\} .
\label{EQNC4}
\ea
If the theory contains massless quarks [i.e., $M=0$ in Eq. (\ref{EQNC4})],
then the partition function $Z^{(0)}_\theta$ comes out to be independent of
$\theta$, being (see also Ref. \cite{Fujikawa79}):
\be
Z^{(0)}_\theta [\eta ,\bar{\eta} ] = Z^{(0)}_{\theta + 2L \alpha}
[\eta', \bar{\eta}'] 
,
\label{EQNC5}
\ee
where $\eta' = e^{-i\alpha\gamma_5}\eta$ and $\bar{\eta}' = \bar{\eta}
e^{-i\alpha\gamma_5}$
(with $\gamma_5 \equiv -i \gamma^0 \gamma^1 \gamma^2 \gamma^3$),
and $L$ is the number of massless flavours. Thus one 
can always reduce $\theta$ to zero by making a global chiral transformation 
on the fermion sources, with $\alpha = -\theta /2L$.
Such a chiral transformation does not change the physics, because all 
Green's function are evaluated in the sourceless limit. Thus all
$\theta$--worlds are physically equivalent to one another, when the theory 
contains massless quarks. In particular, the derivative in (\ref{EQNC4})
vanishes and so $\chi$ is equal to zero.

Let us assume that, to leading order in $1/N_c$, $\chi(k)$ is non--vanishing 
at $k=0$, in the world without quarks. When massless quarks are included, 
$\chi(k)$ must vanish at $k=0$; yet the massless quarks are effects of order
$1/N_c$, since, as we know from 'tHooft's work \cite{tHooft74}, each quark 
loop is suppressed by a factor of $1/N_c$. $\chi(k)$ has the form:
\be
\chi(k) = A_0(k) + A_1(k) + A_2(k) + \ldots ,
\label{EQNC6}
\ee
where $A_0(k)$ is the sum of all diagrams without quark loops, $A_1(k)$ is 
the sum of all diagrams with one quark loop, $A_2(k)$ is the sum of all 
diagrams with two quark loops, and so on.
If one computes $A_0(k)$ in perturbation theory, one finds that it is of 
order $g^4 N^2_c = 1$, since it receives contributions from each of the
$N^2_c-1$ gluon degrees of freedom; moreover each $Q(x)$ takes a factor
$g^2$ which is of order $1/N_c$, since, as 'tHooft showed \cite{tHooft74},
to have a smooth large--$N_c$ limit one should define the coupling constant 
to be $g=g_0/\sqrt{N_c}$, where $g_0$ is kept fixed as $N_c$ becomes large.
Yet $A_1(k)$ is of order $1/N_c$, $A_2(k)$ is of order $1/N^2_c$ and so on:
each successive term is suppressed by an extra factor of $1/N_c$.
If, as we have assumed before, $A_0(k)$ does not vanish at $k=0$, how can 
$\chi(k)$ vanish at $k=0$, since the extra terms in $\chi(k)$ are down by
powers of $1/N_c$? We can answer this question by writing $\chi(k)$ in a
slightly different (yet more explicit) way.

As any other two--point function of gauge invariant operators, $\chi(k)$ can 
be written to the lowest order in $1/N_c$ as a sum over one--hadron poles,
i.e., one--hadron intermediate states (see Ref. \cite{Witten79b}):
\be
\chi(k) = \displaystyle\sum_{\rm glueballs}{a^2_n \over k^2 - M^2_n}
+ \displaystyle\sum_{\rm mesons}{c^2_n /N_c \over k^2 - m^2_n} ,
\label{EQNC7}
\ee
where $M_n$ is the mass of the $n^{\underline{th}}$ glueball state and $m_n$
the mass of the $n^{\underline{th}}$ meson. In this formula $a_n$ is the
matrix element of $Q(0)$ to create the $n^{\underline{th}}$ glueball state:
\be
a_n = \langle 0|Q(0)|n^{\underline{th}}~{\rm glueball} \rangle ,
\label{EQNC8}
\ee
and $c_n/\sqrt{N_c}$ is the matrix element of $Q(0)$ to create the
$n^{\underline{th}}$ meson:
\be
c_n/\sqrt{N_c} = \langle 0|Q(0)|n^{\underline{th}}~{\rm meson} \rangle .
\label{EQNC9}
\ee
From the above--mentioned $1/N_c$--expansion counting rules, we know that
$a_n$ and $c_n$ are both of order one. The first term in (\ref{EQNC7})
(the sum over glueball states) is $A_0(k)$; the second term (the sum over
mesons) is the contribution of quark loops.

We have to explain how the second term, which is suppressed by a factor
$1/N_c$, can cancel the first one. Although this cancellation is impossible 
for a general value of $k$, it can happen at $k=0$ (as we require!) if 
there exists a meson state with a squared mass of order $1/N_c$: since this 
meson couples to $Q(x)$, it must be a pseudoscalar flavour singlet.
We call it the $\eta'$, since the $\eta'$ is the lightest flavour--singlet 
pseudoscalar in nature.
At this point, it is also easy to derive a formula for the mass of the
$\eta'$. The $\eta'$--term is the only meson term in (\ref{EQNC7}) comparable
in magnitude to $A_0(0)$ (both of them being of order $1$), and must by 
itself cancel $A_0(0)$, in order that the total $\chi(0)$ vanishes.
Thus we obtain that:
\be
{c^2_{\eta'} \over N_c m^2_{\eta'}} = A_0(0) .
\label{EQNC10}
\ee
We recall from Eq. (\ref{EQNC9}) that:
\be
{c_{\eta'} \over \sqrt{N_c}} = \langle 0|Q(0)|\eta' \rangle .
\label{EQNC11}
\ee
We can now use the anomaly equation for the $U(1)$ axial current
$J^{(L)}_{5, \mu} = \sum_{i=1}^{L} \bar{q}_i \gamma_\mu \gamma_5
q_i$ ($L$ being the number of massless flavours;
moreover: $\gamma_5 \equiv -i \gamma^0 \gamma^1 \gamma^2 \gamma^3$):
\be
\partial^\mu J^{(L)}_{5, \mu} = 2L Q ,
\label{EQNC12}
\ee
and substitute $Q(0)$ in Eq. (\ref{EQNC11}):
\be
{c_{\eta'} \over \sqrt{N_c}} =
{1 \over 2L} \langle 0|\partial^\mu J^{(L)}_{5, \mu} |\eta' \rangle .
\label{EQNC13}
\ee
We have that $\langle 0|\partial^\mu J^{(L)}_{5, \mu} |\eta' \rangle$ $=$
$-i p^\mu \langle 0|J^{(L)}_{5, \mu} |\eta' \rangle$, and by definition
$\langle 0|J^{(L)}_{5, \mu} |\eta' \rangle$ $=$
$i \sqrt{2L} p_\mu F_{\eta'}$
(with $\sqrt{2L}$ taken out in this way, $F_{\eta'}$ is independent of the
number of flavours, to the lowest order in $1/N_c$).
Combining these formulae, we obtain:
$\langle 0|\partial^\mu J^{(L)}_{5, \mu} |\eta' \rangle$ $=$
$\sqrt{2L}m^2_{\eta'} F_{\eta'}$.
$F_{\eta'}$ is not known experimentally, but to the lowest order in 
$1/N_c$ it is equal to the pion decay constant:
$F_{\eta'} \simeq F_\pi \simeq 93$ MeV.
From all of this we derive that:
\be
{c^2_{\eta'} \over N_c} = {1 \over 4L^2}2L m^4_{\eta'}F^2_\pi =
{1 \over 2L} m^4_{\eta'}F^2_\pi .
\label{EQNC14}
\ee
Substituting this into Eq. (\ref{EQNC10}), we finally obtain the formula for
the squared mass of the $\eta'$:
\be
m^2_{\eta'} = {2L A \over F^2_\pi} ,
\label{EQNC15}
\ee
where $A \equiv A_0(0)$ is just the topological susceptibility in a QCD
without quarks, i.e., in a pure Yang--Mills theory.

Let us make some comments about formula (\ref{EQNC15}).
From the $1/N_c$--expansion theory, it is known that $F_\pi$ is of order
${\cal O}(\sqrt{N_c})$ (see Ref. \cite{Witten79b}:
in fact the current--current correlation function,
which is of order $N_c$, has a term $F^2_\pi /k^2$). Since $A \equiv A_0(0)$
is of order ${\cal O}(1)$ in $N_c$, from Eq. (\ref{EQNC15})
we derive that $m^2_{\eta'}$ is of order ${\cal O}(1/N_c)$.
This just means that the mass of the $\eta'$ vanishes in the limit
$N_c \to \infty$. In the same limit also the anomaly vanishes, since we 
have that:
\be
Q(x) = {g^2_0 \over 64\pi^2 N_c}\varepsilon^{\mu\nu\rho\sigma}
F^a_{\mu\nu}F^a_{\rho\sigma} = {\cal O} \left( {1 \over N_c} \right) .
\label{EQNC16}
\ee
The fact that the anomaly vanishes when $N_c \to \infty$ is perfectly 
natural, since the anomaly comes from a diagram with one quark loop, and 
quark loops are of order $1/N_c$. At $N_c = \infty$, the $U(1)$ current is 
conserved and (unless there is a physical axial symmetry) there must be a 
massless $U(1)$ Goldstone boson: it is the $\eta'$. At finite (but large) 
$N_c$, it receives a mass of order $1/N_c$: also this seems to be natural, 
because the squared mass of an approximate Goldstone boson is linear in the 
symmetry breaking parameter. The quantity $A$ has been determined by Monte
Carlo simulations in lattice pure--gauge QCD, obtaining a result of the right
order of magnitude (see, e.g., Refs. \cite{Teper88,Campostrini-SU(3)} and
\cite{Alles-et-al.97}):
\be
A \simeq (150 \div 200 ~{\rm MeV})^4 .
\label{EQNC17}
\ee

\newsection{Veneziano's mechanism}

\noindent
Just after Witten's proposal that the $U(1)$ problem could be solved in 
$1/N_c$--expanded QCD, G. Veneziano showed \cite{Veneziano79} that this picture 
comes out to be automatically consistent with expected 
$\theta$--dependences and anomalous Ward Identities, if a (modified)
Kogut--Susskind mechanism is used. This is what we call {\it ``Veneziano's 
mechanism''}, which we now describe in its main aspects.

We must first formulate precisely the $U(1)$ problem in $1/N_c$ language 
and discuss which is the relevant large--$N_c$ limit to be taken
into account.
Let us consider a QCD Lagrangian with an $SU(N_c)$ colour gauge group and 
$N_f$ quark flavours. $L$ of these are taken to be light quarks with common 
mass $m \ll \Lambda_{QCD}$, while the remaining ones have masses greater 
than $\Lambda_{QCD}$ ($\Lambda_{QCD}$ is, as usual, the renormalization 
group invariant scale of QCD: in the $\overline{MS}$ renormalization scheme,
for example, $\Lambda_{QCD} \sim 0.5$ GeV).

We consider the {\it Topological Expansion} (TE) of QCD discussed in
Ref. \cite{Veneziano76} and obtained by taking $N_c,N_f \to \infty$ with a 
fixed ratio and with fixed $g^2 N_c$. If at the leading (planar) level we 
have both confinement and the Nambu--Goldstone (NG) mechanism, then there 
is an $L^2$--plet of light NG pseudoscalars of mass:
\ba
m^2_{NS} = F^{-2}_{NS} \langle -2m \bar{q}_a q_a \rangle \equiv F^{-2}_{NS}
\langle \Delta_2 \rangle, \nonumber \\
a = 1,2,...,L,  ~~~ {\rm no~ sum ~over}~ a,
\label{EQND1}
\ea
where $NS$ stands for ``{\it non--singlet}'' of $SU(L)_{flavour}$;
$F_{NS}$ is the analogous of $F_\pi$; both $F^2_{NS}$ and
$\langle \bar{q}q\rangle$ are of order $N_c$
\cite{tHooft74,Veneziano76,Witten79a,Witten79b},
so that $m^2_{NS} \propto m\Lambda_{QCD}$.
At this planar level $m_{NS}$ is also the mass of the $SU(L)$ singlet, 
because there are no OZI violating $q\bar{q}$ annihilation diagrams at 
leading order in $1/N_c$: the $L^2$ physical mesons are diagonal in the
$q\bar{q}$ basis and all degenerate.
Another consequence of planarity is that decay couplings ($F_\pi ,F_K ,
\ldots $) are all related. We find that:
\ba
\langle NS{\rm -pseudoscalar},a|A^b_\mu (0)|0 \rangle =
-i\delta_{ab}q_\mu F_{NS} ,
\nonumber \\
\langle S{\rm -pseudoscalar}|J^{(L)}_{5, \mu} (0)|0 \rangle =
-iq_\mu \sqrt{2L}F_{NS} ,
\label{EQND2}
\ea
where the $A^a_\mu = \bar{q}\gamma_\mu \gamma_5 {1 \over 2}\lambda^a q$  
are the $SU(L)$ axial currents and
$J^{(L)}_{5, \mu} = \sum_{i=1}^{L} \bar{q}_i \gamma_\mu \gamma_5
q_i$ is the $U(1)$ axial current.
All this happens at the planar level. When going to the next--to--leading 
(non planar) diagrams, we must find that the $L^2$ degeneracy is broken 
down to an $L^2 -1$ degeneracy of NG bosons together with a heavier $SU(L)$ 
singlet of finite mass, even in the chiral limit $m ~(m^2_{NS}) \to 0$.
Since we are taking the limit of light quark masses ($m/\Lambda_{QCD} \to 
0$), it turns out that, to have a smooth limit, it is better to consider 
the TE expansion of QCD, rather than the simple $N_c \to \infty$ limit of 
'tHooft. In this TE expansion one keeps $L/N_c$ fixed with inclusion of an 
arbitrary number of quark loops at each order, and then one lets (for
simplicity) $L/N_c \to 0$.
More precisely, the limit that one considers is the following:
\be
m^2_{NS} \sim m\Lambda_{QCD} \ll {L \over N_c}\Lambda^2_{QCD} 
\ll \Lambda^2_{QCD} .
\label{EQND3}
\ee
In this limit one can saturate algebrically the Ward Identities (WI's) of the 
theory. The relevant WI's of QCD have the form \cite{Crewther77}:
\be
\int d^4 x \langle T\partial^\mu A^a_\mu (x) \partial^\nu A^b_\nu (0)\rangle
= -i \delta _{ab} \langle \Delta_2 \rangle ,
\label{EQND4}
\ee
\be
2L\int d^4 x \langle T Q(x) D_{(L)}(0) \rangle =
-2iL \langle \Delta_2 \rangle
- \int d^4 x \langle T D_{(L)}(x) D_{(L)}(0) \rangle ,
\label{EQND5}
\ee
\be
4L^2\int d^4 x \langle T Q(x) Q(0) \rangle = 2iL \langle \Delta_2 \rangle
+ \int d^4 x \langle T D_{(L)}(x) D_{(L)}(0) \rangle .
\label{EQND6}
\ee
Here $D_{(L)} = 2im\sum_{i=1}^L \bar{q}_i \gamma_5 q_i$ is the
mass term appearing in the divergence of the $U(1)$ axial current:
$\partial^\mu J^{(L)}_{5, \mu}(x) = D_{(L)}(x) + 2L Q(x)$; and 
$\Delta_2 = -2m\bar{q}_a q_a$ (no sum over $a=1,2, \ldots ,L$).
Eq. (\ref{EQND4}) is the usual anomaly--free WI for the $SU(L)$--non-singlet 
currents: it implies the existence of NG bosons of mass $m_{NS}$
given by (\ref{EQND1}).

We will concentrate on the two anomalous WI's, (\ref{EQND5}) and (\ref{EQND6}).
Let us assume, as proposed by Witten \cite{Witten79a} (see Section 4),
that the left--hand side of Eq. (\ref{EQND6}) is non--zero at the
leading order in $1/N_c$ of pure Yang--Mills (YM) theory. To obtain this, we 
introduce an axial four--vector ghost of propagator $-ig_{\mu\nu}/q^2$, 
coupled to the current $K_\mu$ of Eq. (\ref{EQNB2}) with strength 
$i\lambda_{YM}(q^2)$, where $\lambda_{YM} = {\cal O}(1)$.
Therefore, in pure YM and large $N_c$, the left--hand side of
Eq. (\ref{EQND6}) may be written as:
\be
4L^2 (i\lambda_{YM}(q^2))^2(iq_\mu)(-iq_\nu){-ig_{\mu\nu} \over q^2}
\mid _{q^2 \to 0} =
4iL^2\lambda^2_{YM}(0) = {\cal O}(L^2) .
\label{EQND7}
\ee
Instead, the right--hand side is of order ${\cal O}(mLN_c)$ (in the absence of
$SU(L)$ singlets of squared mass ${\cal O}(m)$). To eliminate this discrepancy,
we have to make a resummation of quark loops: in fact, even if they are 
suppressed by factors of $(L/N_c)^n$, they contain intermediate states of 
squared mass of order ${\cal O}(m)$.
Thus we also assume that the ghost couples to the (planar) $SU(L)$ singlet
(which we will denote by $\eta_L$) with strength
$i\sqrt{L/N_c}\lambda_\eta (q^2)q_\mu$ (with $\lambda_\eta$ again 
independent of $L$, $N_c$ and $m$). Then we can sum all quark loops in the 
leading non--planar (cylinder) topology, obtaining the following expression 
for the right--hand side (RHS) of Eq. (\ref{EQND6}):
\ba
\lefteqn{
4iL^2\lambda^2_{YM}(q^2) \left[ 1+{L \over N_c} \lambda^2_\eta (q^2)
{1 \over q^2 - m^2_{NS}} + \ldots \right] _{q^2 \to 0} } \nonumber \\
& & = {4iL^2 \lambda^2_{YM}(q^2) (q^2 - m^2_{NS}) \over
q^2 - m^2_{NS} - {L \over N_c}\lambda^2_\eta (q^2)} 
\mid _{q^2 \to 0} 
= {4iL^2\lambda^2_{YM}(0)m^2_{NS} \over m^2_{NS} + {L \over N_c}
\lambda^2_\eta (0)} .
\label{EQND8}
\ea
As a consequence of the resummation, now the RHS of Eq. (\ref{EQND6}) is of
order ${\cal O}(mLN_c)$, as required by the WI (remember that $m^2_{NS} =
{\cal O}(m)$). When considering the two--point function of $Q(x)$ at
four--momentum $q$, we find that:
\be
4L^2\int d^4x e^{iqx} \langle T Q(x)Q(0) \rangle = 
{4iL^2 \lambda^2_{YM}(q^2) (q^2 - m^2_{NS}) \over
q^2 - (m^2_{NS} + {L \over N_c}\lambda^2_\eta (q^2))} .
\label{EQND9}
\ee
Using the expansion (\ref{EQNC7}) given in Section 4, one
immediately derives from Eq. (\ref{EQND9}) that the $\eta_L$--mass $m_S$
is given by:
\be
m^2_{\eta_L} \equiv m^2_S = m^2_{NS} + {L \over N_c}\lambda^2_\eta
(m^2_{S}) .
\label{EQND10}
\ee
Furthermore, by going on the $\eta_L$--pole and using Eq. (\ref{EQND2}),
one obtains that:
\be
\langle 0|Q(0)|\eta_L \rangle = {F_{NS} \over \sqrt{2L}}m^2_S
= \lambda_{YM}(m^2_{S}) \lambda_\eta (m^2_S) \sqrt{{L \over N_c}} .
\label{EQND11}
\ee
Combining Eqs. (\ref{EQND10}) and (\ref{EQND11}), we get, in the chiral
limit $m~(m^2_{NS}) \to 0$:
\be
m^2_S = {2L\lambda^2_{YM} (m^2_S) \over F^2_{NS}} .
\label{EQND12}
\ee
We can assume that the $q$--dependences of $\lambda_{YM}$ and 
$\lambda_\eta$ are really negligible, since the range of $q^2$ to be 
considered is $0 \le q^2 \le (L/N_c)\Lambda^2_{QCD} \ll \Lambda^2_{QCD}$ 
and variations in this range are non--leading in $L/N_c$, if we assume
no massless glueballs. (A lattice confirmation of the $q$--independence of
$\lambda_{YM}$ was given in Refs. \cite{Briganti-et-al.91} and \cite{EM1992b}.)
Therefore we may substitute everywhere $\lambda_{YM}(m^2_S)$ and
$\lambda_\eta (m^2_S)$ with their values calculated at $q^2=0$:
$\lambda_{YM}(0)$ and $\lambda_\eta (0)$.
In this way, Eq. (\ref{EQND12}) acquires exactly the form of
Eq. (\ref{EQNC15}) derived by Witten:
\be
m^2_S = {2L\lambda^2_{YM} (0) \over F^2_{NS}} ,
\label{EQND13}
\ee
$\lambda^2_{YM}(0)$ being precisely the YM topological susceptibility
(see Eq. (\ref{EQND7})).
Using Eqs. (\ref{EQND8}), (\ref{EQND10}) and (\ref{EQND13}), one can
easily verify that the WI (\ref{EQND6}) is saturated (i.e., its left--hand
side and its right--hand side turn out to be equal already at this level of
approximation):
\be
4iLN_c m^2_{NS} \left( {\lambda_{YM}(0) \over \lambda_\eta(0)} \right)^2
= 2iLm^2_{NS} F^2_{NS} .
\label{EQND14}
\ee
And the same happens also for the other two WI's. In the above--described
picture we have considered an {\it ideal world} where we have $N_c$ colours and
$L$ light flavours of common mass $m \ll \Lambda_{QCD}$. 
In the real world we have $N_c = 3$ colours and $L=3$ light flavours, with 
different masses: $m_u,~m_d,~m_s \ll \Lambda_{QCD}$ (for $m_s$ this relation 
is less strong that for $m_u$ and $m_d$). Nevertheless one can use the above
results (in particular Eq. (\ref{EQND10})) to derive 
the squared mass matrix for the pseudoscalar nonet ($\pi ,K ,\eta ,\eta'$).
What one finds at the end is the following very interesting formula for the 
$\eta'$--mass:
\be
{6A \over F^2_\pi} = m^2_{\eta'} + m^2_\eta -2 m^2_K ,
\label{EQND15}
\ee
where $A=\lambda^2_{YM}(0)$ is the YM topological susceptibility. Inserting 
here the experimental values of the meson masses and of the pion decay 
constant ($F_\pi \simeq 93$ MeV), one derives that $A$ is of order
$A \simeq (180~{\rm MeV})^4$. Lattice calculations have confirmed such an order
of magnitude for $A$ (see, e.g., Refs. \cite{Teper88,Campostrini-SU(3)} and
\cite{Alles-et-al.97}).

\newsection{The phase transitions of QCD}

\noindent
All the above refers to the theory at $T=0$.
In the following of the paper, we discuss the role of the $U(1)$ axial
symmetry for the phase structure of QCD at finite temperature.
One expects that, above a certain critical temperature, also the $U(1)$ axial
symmetry will be restored. We will try to see if this transition has (or has
not) anything to do with the usual chiral transition.
Let us define the following temperatures:
\begin{itemize}
\item{} $T_{ch}$: the temperature at which the chiral condensate
$\langle \bar{q} q \rangle$ goes to zero. The chiral symmetry
$SU(L) \otimes SU(L)$ is spontaneously broken below $T_{ch}$ and
it is restored above $T_{ch}$.
\item{} $T_{U(1)}$: the temperature at which the $U(1)$ axial symmetry
is (approximately) restored.
If $\langle \bar{q} q \rangle \ne 0$ also the $U(1)$ axial symmetry is
broken, i.e., the chiral condensate is an order parameter also for
the $U(1)$ axial symmetry. Therefore we must have: $T_{U(1)} \ge T_{ch}$.
\item{} $T_\chi$: the temperature at which the pure--gauge topological
susceptibility $A$ (approximately) drops to zero. Present lattice results
indicate that $T_\chi \ge T_{ch}$ \cite{Alles-et-al.97}.
Moreover, the Witten--Veneziano mechanism implies that $T_{U(1)} \ge T_\chi$.
\end{itemize}
The following scenario, that we will call ``\underline{SCENARIO 1}'',
in which $T_\chi < T_{ch}$, is, therefore, immediately ruled out.
In this case, in the range of temperatures between $T_\chi$ and $T_{ch}$
the $U(1)$ axial symmetry is still broken by the chiral condensate, but the
anomaly effects are absent. In other words, in this range of temperatures
the $U(1)$ axial symmetry is spontaneously broken ({\it \`a la} Goldstone)
and the $\eta'$ is the corresponding NG boson, i.e., it is massless in the
chiral limit $\sup(m_i) \to 0$, or, at least, as light as the pion $\pi$,
when including the quark masses.
This scenario was first discussed (and indeed really supported!) in
Ref. \cite{Pisarski-Wilczek84}.
It is known that the $U(1)$ chiral anomaly effects are related with
instantons \cite{tHooft76}. It is also known that at high temperature $T$
the instanton--induced amplitudes are suppressed by the so--called
``Pisarski--Yaffe suppression factor'' \cite{Gross-Pisarski-Yaffe81},
due to the Debye--type screening:
\be
d{\cal N}_{inst}(T) \sim d{\cal N}_{inst}(T=0) \cdot
\exp \left[ -\pi^2 \rho^2 T^2 \left( {2N_c + L \over 3} \right) \right] ,
\label{eqn4}
\ee
$\rho$ being the instanton radius.
The argument of Pisarski and Wilczek in Ref. \cite{Pisarski-Wilczek84} was
the following: {``If instantons themselves are the primary 
chiral--symmetry--breaking mechanism, then it is very difficult to
imagine the unsuppressed $U(1)_A$ amplitude at $T_{ch}$.''}
So, what was wrong in their argument?
The problem is that Eq. (\ref{eqn4}) can be applied only in the quark--gluon
plasma phase, since the Debye screening is absent below $T_{ch}$.
Indeed, Eq. (\ref{eqn4}) is applicable only for $T \ge 2T_{ch}$ and
one finds instanton suppression by at least two orders of magnitude
at $T \simeq 2T_{ch}$ (see Ref. \cite{Shuryak94} and references therein).
Moreover, the qualitative picture of instanton--driven chiral symmetry
restoration which is nowadays accepted has significantly changed since
the days of Ref. \cite{Pisarski-Wilczek84}.
It is now believed (see Ref. \cite{Shuryak94} and references therein)
that the suppression of instantons is not the only way to ``kill'' the
quark condensate. Not only the number of instantons is important, but
also their relative positions and orientations. Above $T_{ch}$,
instantons and anti-instantons can be rearranged into some finite
clusters with zero topological charge, such as well--formed
``instanton--anti-instanton molecules''.

Therefore, we are left essentially with the two following scenarios.\\
\underline{SCENARIO 2}: $T_{ch} \le T_{U(1)}$, with
$T_{ch} \sim T_\chi \sim T_{U(1)}$.
If $T_{ch} = T_\chi = T_{U(1)}$, then, in the case of $L=2$ light flavours,
the restored symmetry across the transition is $U(1)_A \otimes SU(2)_L \otimes
SU(2)_R \sim O(2) \otimes O(4)$, which may yield a first--order phase
transition (see, for example, Ref. \cite{Kharzeev-et-al.98}).\\
\underline{SCENARIO 3}: $T_{ch} \ll T_{U(1)}$, that is, the complete
$U(L)_L \otimes U(L)_R$ chiral symmetry is restored only well inside the
quark--gluon plasma domain.
In the case of $L=2$ light flavours, we then have at $T=T_{ch}$ the
restoration of $SU(2)_L \otimes SU(2)_R \sim O(4)$.
Therefore, we can have a second--order phase transition with the
$O(4)$ critical exponents. $L=2$ QCD at $T \simeq T_{ch}$ and the $O(4)$
spin system should belong to the same universality class.
An effective Lagrangian describing the softest modes is essentially
the Gell-Mann--Levy linear sigma model, the same as for the $O(4)$
spin systems (see Ref. \cite{Pisarski-Wilczek84}).
If this scenario is true, one should find the $O(4)$ critical indices
for the quark condensate and the specific heat:
$\langle \bar{q} q \rangle \sim |(T - T_{ch})/T_{ch}|^{0.38 \pm 0.01}$,
and $C(T) \sim |(T - T_{ch})/T_{ch}|^{0.19 \pm 0.06}$.
Present lattice data partially support these results.

\newsection{The $U(1)$ chiral order parameter}

\noindent
We make the assumption (discussed in the Introduction and in the previous
Section) that the $U(1)$ chiral symmetry is broken independently from the
$SU(L) \otimes SU(L)$ symmetry. The usual chiral order parameter
$\langle \bar{q} q \rangle$ is an order parameter both for
$SU(L) \otimes SU(L)$ and for $U(1)_A$: when it is different from zero,
$SU(L) \otimes SU(L)$ is broken down to $SU(L)_V$ (``$V$'' stands for
{\it ``vectorial''}) and also $U(1)_A$ is broken.
Thus we need another quantity which could be an order parameter only for 
the $U(1)$ chiral symmetry \cite{EM1994a,EM1994b,EM1994c,EM1995a}.
The most simple quantity of this kind was found by 'tHooft in 1976 (see 
\cite{tHooft76}) studying the effective Lagrangian of instantons (so 
that, for historical reasons, we will call this quantity ``${\cal L}_{eff}$'').
For a theory with $L$ light quark flavours (of mass $m_i \ll \Lambda_{QCD}$;
$i=1,\ldots ,L$), it is a $2L$--fermion interaction that has the chiral 
transformation properties of:
\be
{\cal L}_{eff}^{(L)} \sim \displaystyle {\det_{st}} \left[ \bar{q}_s
\left( {1+\gamma_5 \over 2} \right) q_t \right]  + {\rm h.c.} =
\displaystyle {\det_{st}} \left( \bar{q}_{sR} q_{tL} \right) +
\displaystyle {\det_{st}} \left( \bar{q}_{sL} q_{tR} \right) ,
\label{EQN2.1}
\ee
where $s,t = 1, \ldots ,L$ are flavour indices, but the colour indices are 
arranged in a more general way  (see below); as usual,
$q_L \equiv {1 \over 2}(1 + \gamma_5)q$ and
$q_R \equiv {1 \over 2}(1 - \gamma_5)q$,
with $\gamma_5 \equiv -i \gamma^0 \gamma^1 \gamma^2 \gamma^3$,
denote respectively the {\it left--handed} and the {\it right--handed} quark
fields.
Since under chiral $U(L) \otimes U(L)$ transformations the quark fields
transform as follows:
\be
q_L \to V_L q_L ~~~,~~~q_R \to V_R q_R,
\label{EQN2.2}
\ee
where $V_L$ and $V_R$ are arbitrary $L \times L$ unitary matrices, 
we immediately derive the transformation property of
${\cal L}_{eff}^{(L)}$ under $U(L) \otimes U(L)$:
\be
U(L) \otimes U(L):~~~ {\cal L}_{eff}^{(L)} \to \det(V_L)\det(V_R)^*
\displaystyle {\det_{st}} \left( \bar{q}_{sR} q_{tL} \right) + {\rm h.c.} .
\label{EQN2.3}
\ee
This just means that ${\cal L}_{eff}^{(L)}$ is invariant under
$SU(L) \otimes SU(L) \otimes U(1)_V$, while it is not invariant under $U(1)_A$:
\be
U(1)_A:~~~ {\cal L}_{eff}^{(L)} \to e^{-i 2L \alpha}
\displaystyle {\det_{st}} \left( \bar{q}_{sR} q_{tL} \right) + {\rm h.c.} ,
\label{EQN2.4}
\ee
where $\alpha$ is the parameter of the $U(1)$ chiral transformation 
($U(1)_A$: $q \to e^{-i\alpha\gamma_5}q$, i.e., $q_L \to e^{-i\alpha}q_L$
and $q_R \to e^{i\alpha}q_R$).

As an example let us consider the most simple case, that is $L=2$, but with 
a general colour group $SU(N_c)$.
It is not hard to find (using the Fierz relations both for the spinorial 
matrices and the $SU(N_c)$ generators in their fundamental representation) 
that the most general colour--singlet, hermitian and P--invariant 
local quantity (without derivatives)
which has the required chiral transformation properties is just the following  
four--fermion local operator:
\be
{\cal L}_{eff}^{(L=2)}(\alpha_0 ,\beta_0) = 
F_{bd}^{ac}(\alpha_0 ,\beta_0) \epsilon^{st}
\left( \bar{q}_{1R}^{a}q_{sL}^{b} \cdot
\bar{q}_{2R}^{c}q_{tL}^{d} +
\bar{q}_{1L}^{a}q_{sR}^{b} \cdot
\bar{q}_{2L}^{c}q_{tR}^{d} \right) ,
\label{EQN2.5}
\ee
where the ``colour tensor'' $F_{bd}^{ac}(\alpha_0 ,\beta_0)$ is given by:
\be
F_{bd}^{ac}(\alpha_0 ,\beta_0) = 
\alpha_0 \delta_b^a \delta_d^c + \beta_0 \delta_d^a \delta_b^c,
\label{EQN2.6}
\ee
$\alpha_0$ and $\beta_0$ being arbitrary real parameters. 
In Eq. (\ref{EQN2.5}), $a,b,c,d = 1, \ldots ,N_c$ are colour indices; $s$ and
$t$ are flavour indices and $\epsilon^{st}=-\epsilon^{ts}$,
$\epsilon^{12}=1$. Dirac indices are contracted between the first and the 
second fermion field and also between the third and the fourth one.
Note that if we choose $\alpha_0 = N_c$ and $\beta_0 = -1$, 
${\cal L}_{eff}^{(L=2)}(\alpha_0 ,\beta_0)$ 
just becomes (up to a proportionality constant) the effective Lagrangian 
found by 'tHooft in \cite{tHooft76} to describe the instantons.

Now, to obtain an order parameter for the $U(1)$ chiral symmetry, one can 
simply take the vacuum expectation value of 
${\cal L}_{eff}^{(L=2)}(\alpha_0 ,\beta_0)$:
\be
C_{U(1)}^{(L=2)}(\alpha_0 ,\beta_0) \equiv 
\langle {\cal L}_{eff}^{(L=2)}(\alpha_0 ,\beta_0) \rangle .
\label{EQN2.7}
\ee
The arbitrarity in the choice of $\alpha_0$ and $\beta_0$ (indeed of only 
one of them, since only their ratio is relevant) can be removed if we 
require that the new $U(1)$ chiral condensate is ``independent'', in a 
sense which will be explained below, of the usual chiral condensate
$\langle \bar{q} q \rangle$.
As it was pointed out by Shifman, Vainshtein and Zakharov in 
\cite{Shifman-Vainshtein-Zakharov79}, a matrix element of the form 
$\langle \bar{q} \Gamma_1 q \cdot \bar{q} \Gamma_2 q \rangle$ 
has a big contribution proportional to the square 
of the vacuum expectation value (v.e.v.) of $\bar{q} q$. This 
contribution corresponds to retaining the vacuum intermediate state in all 
the channels and neglecting the contributions of all the other states; we 
call this contribution the {\it ``disconnected part''} of the original matrix 
element:
\be
\langle \bar{q} \Gamma_1 q \cdot \bar{q} \Gamma_2 q \rangle_{disc.} 
= {1 \over N^2} \left[ (\Tr\Gamma_1 \cdot \Tr\Gamma_2) - 
\Tr(\Gamma_1 \Gamma_2) \right] \langle \bar{q} q \rangle^2,
\label{EQN2.8}
\ee
where the normalization factor $N$ is defined as:
\be
\langle \bar{q}_A q_B \rangle =
{\delta_{AB} \over N} \langle \bar{q} q \rangle,
\label{EQN2.9}
\ee
(that is $N = \delta_{AA}$; $\bar{q} q = \sum_A{\bar{q}_A q_A}$) 
and the subscripts $A,B$ include spin, colour and flavour.
When considering the operator ${\cal L}_{eff}^{(L=2)}(\alpha_0 ,\beta_0)$ 
defined in Eqs. (\ref{EQN2.5}) and (\ref{EQN2.6}), we find the following
expression for its {\it disconnected} part:
\be
\langle {\cal L}^{(L=2)}_{eff}(\alpha_0 ,\beta_0) \rangle_{disc.} =
{1 \over 16 N_c} [N_c (2\alpha_0 + \beta_0) + (\alpha_0 + 2\beta_0)]  
\langle \bar{q} q \rangle^2 ,
\label{EQN2.10}
\ee
(where: $\langle \bar{q} q \rangle = \langle \bar{u} u \rangle
+ \langle \bar{d} d \rangle$).
From this last equation we immediately see that the {\it disconnected}
part of the condensate $C_{U(1)}^{(L=2)}(\alpha_0 ,\beta_0)$ vanishes
with the following particular choice of the coefficients $\alpha_0$ and
$\beta_0$ (only their ratio is really relevant):
\be
{\beta_0 \over \alpha_0} = -{2N_c + 1 \over N_c + 2}.
\label{EQN2.11}
\ee 
In other words, the condensate Eq. (\ref{EQN2.7}) with $\alpha_0$ and $\beta_0$
satisfying the constraint Eq. (\ref{EQN2.11}) does not take contributions from
the usual chiral condensate $\langle \bar{q} q \rangle$.
To summarize, a good choice for a $U(1)$ chiral condensate which is really
``independent'' of the usual chiral condensate $\langle \bar{q} q \rangle$
is the following one (apart from an irrelevant multiplicative constant):
\be
C_{U(1)}^{(L=2)} = 
\langle ( \delta_b^a \delta_d^c  -{2N_c + 1 \over N_c + 2} 
\delta_d^a \delta_b^c ) 
\epsilon^{st}
\left( \bar{q}_{1R}^{a}q_{sL}^{b} \cdot
\bar{q}_{2R}^{c}q_{tL}^{d} +
\bar{q}_{1L}^{a}q_{sR}^{b} \cdot
\bar{q}_{2L}^{c}q_{tR}^{d} \right) \rangle .
\label{EQN2.12}
\ee
As a remark, we observe that the condensate $C_{U(1)}^{(L=2)}$ so defined
turns out to be of order ${\cal O}(g^2 N_c^2) = {\cal O}(N_c)$ in the
large--$N_c$ expansion (this will be derived also in Section 8 by simply
requiring that the $1/N_c$ expansion of the relevant QCD Ward Identities
remains well defined when including this new condensate \cite{EM1994b}).
In the particular (yet physical!) case $N_c =3$, the condensate
Eq. (\ref{EQN2.12}) becomes:
\be
C_{U(1)}^{(L=2)} = 
\langle ( \delta_b^a \delta_d^c  -{7 \over 5} \delta_d^a \delta_b^c )
\epsilon^{st} \left( \bar{q}_{1R}^{a}q_{sL}^{b} \cdot
\bar{q}_{2R}^{c}q_{tL}^{d} + \bar{q}_{1L}^{a}q_{sR}^{b} \cdot
\bar{q}_{2L}^{c}q_{tR}^{d} \right) \rangle .
\label{EQN2.13}
\ee
So far we have considered the most simple case $L=2$. However, this
procedure can be easily generalized to every $L$, and we can take as
an order parameter for the $U(1)$ chiral symmetry:
\be
C^{(L)}_{U(1)} = \langle {\cal L}_{eff}^{(L)} \rangle .
\label{EQN2.17}
\ee
As we have done in the case $L=2$, the colour indices may be arranged in such
a way that the $U(1)$ chiral condensate does not take contributions from the
usual chiral condensate $\langle \bar{q} q \rangle$: as a consenquence of
this, the new condensate will be of order ${\cal O}(g^{2L - 2} N_c^L) =
{\cal O}(N_c)$ in the large--$N_c$ expansion.
The case of the {\it real world} is that in 
which $L=3$ (three light flavours, $u,d$ and $s$, with masses $m_u, m_d, 
m_s$ small compared to the QCD mass--scale $\Lambda_{QCD}$). In the next
Sections we will try to derive which are the physical consequences 
of assuming that $C_{U(1)}^{(L)} \ne 0$ up to a temperature $T_{U(1)}$
far above the usual chiral  critical temperature $T_{ch}$.
For doing this, we will analyse the relevant QCD Ward Identities by making
an expansion in $1/N_c$ and a chiral expansion in the light quark masses.
In Section 9 we will see how to derive a new chiral effective Lagrangian which
also incorporates the $U(1)$ chiral order parameter.

\newsection{Saturation of Ward Identities}

\noindent
In this Section we make an analysis of the relevant QCD Ward Identities (WI's).
Let us consider a set of {\it ideal worlds} described by the usual QCD
Lagrangian with an $SU(N_c)$ colour gauge group and $N_f$ quark flavours,
where $N_c$ and $N_f$ are considered as variables. Among these $N_f$ quarks,
$L$ are taken to be light quarks with mass $m_l \ll \Lambda_{QCD}$,
$\Lambda_{QCD}$ being the renormalization--group invariant scale of QCD.
The saturation of the relevant QCD Ward Identities will be obtained in a
$1/N_c$ expansion, $N_c$ being the number of colours
\cite{tHooft74,Veneziano76}.

When deriving the relevant WI's as in Refs. \cite{Crewther79} and
\cite{Veneziano79}, one only uses the canonical (anti--)commutation
relations at equal times:
\be
\{ q_i(x) , q^\dagger _j (x^\prime ) \}^{+}_{x^0 = x^{\prime 0}}
= \delta_{ij} \delta ( {\bf x} - {\bf x}^\prime ) ,
\label{EQN10.7}
\ee
which do not absolutely depend on the temperature $T$ of the physical 
system: they are a fundamental law of quantum mechanics. For this reason 
the relevant WI's at $T \ne 0$ are simply derived from those at $T=0$ by 
substituting everywhere the vacuum expectation values with the usual thermal 
average over the Gibbs ensemble ($\beta =1/T$):
\be
\langle 0|O|0 \rangle \to {\Tr[e^{-\beta H}O] \over \Tr[e^{-\beta H}]} 
={\displaystyle\sum_n{e^{-\beta E_n} \langle n|O|n \rangle } \over
\displaystyle\sum_n{e^{-\beta E_n}} } \equiv \langle O \rangle ,
\label{EQN10.8}
\ee
$H$ being the total hamiltonian of the system, $E_n$ and $|n \rangle$ its 
eigenvalues and eigenstates ($H|n \rangle =E_n|n \rangle$).

In the case of the $SU(L) \otimes SU(L)$ chiral symmetry, one
immediately verifies that:
\be
\langle [Q_A^a(x_0), i\bar{q} \gamma_5 T^b q(y)]_{x_0 = y_0} \rangle
= -i \delta_{ab} {1 \over L} \langle \bar{q} q \rangle ,
\label{comm-su}
\ee
where $Q_A^a(x_0)$ is the charge operator for the $SU(L)$ chiral symmetry
(i.e., $Q_A^a(x_0) \equiv \int{d^3 {\bf x} A^a_0 ({\bf x},x_0)}$,
where $A^a_\mu = \bar{q}\gamma_\mu \gamma_5 T^a q$ is the $SU(L)$ axial
current). From Eq. (\ref{comm-su}) one derives the following WI:
\be
\int d^4 x \langle T\partial^\mu A^a_\mu (x) 
i\bar{q} \gamma_5 T^b q(0) \rangle
= i \delta_{ab} {1 \over L} \langle \bar{q} q \rangle ,
\label{eqn6}
\ee
If $\langle \bar{q} q \rangle \ne 0$ (in the chiral limit
$\sup(m_i) \to 0$), the anomaly--free WI (\ref{eqn6}) implies the existence
of $L^2-1$ non--singlet NG bosons, interpolated by the hermitian fields
$O_b = i \bar{q} \gamma_5 T^b q$.
Similarly, in the case of the $U(1)$ axial symmetry, one finds that:
\be
\langle [Q_5^{(L)}(x_0), i\bar{q} \gamma_5 q(y)]_{x_0 = y_0} \rangle
= -2i \langle \bar{q} q \rangle ,
\label{comm-u1}
\ee
where $Q_5^{(L)}(x_0)$ is the charge operator for the $U(1)$ chiral symmetry
(i.e., $Q_5^{(L)}(x_0) \equiv \int{d^3 {\bf x} J^{(L)}_{5, 0}
({\bf x},x_0)}$, where $J^{(L)}_{5, \mu}= {\bar{q} \gamma_\mu \gamma_5 q}$
is the usual $U(1)$ axial current).
From Eq. (\ref{comm-u1}) one derives the following WI:
\be
\int d^4x \langle T\partial^\mu J^{(L)}_{5, \mu}(x) i\bar{q} \gamma_5 q(0)
\rangle = 2i \langle \bar{q} q \rangle .
\label{eqn7}
\ee
But this is not the whole story! One also derives that:
\be
[Q_5^{(L)}(x_0) , O_P^{(L)}(y)]_{x_0 = y_0} =
-2Li \cdot {\cal L}_{eff}^{(L)}(y) ,
\label{EQN10.4}
\ee
where ${\cal L}_{eff}^{(L)} \sim {\det}(\bar{q}_{sR}q_{tL})
+ {\det}(\bar{q}_{sL}q_{tR})$ is the $2L$--fermion operator
discussed in the previous Section and the hermitian field $O_P^{(L)}$ is so
defined:
\be
O_P^{(L)} \sim i[ \displaystyle{\det_{st}}(\bar{q}_{sR}q_{tL})
- \displaystyle{\det_{st}}(\bar{q}_{sL}q_{tR}) ].
\label{O_P}
\ee
From Eq. (\ref{EQN10.4}) one derives the following WI:
\be
\int d^4x \langle T\partial^\mu J^{(L)}_{5, \mu}(x) O_P^{(L)}(0) \rangle = 
2Li \langle {\cal L}_{eff}^{(L)}(0) \rangle .
\label{EQN10.5}
\ee
Remembering that $\partial^\mu J^{(L)}_{5, \mu} = D_{(L)}+2LQ$, where
$D_{(L)}=2i\sum_{l=1}^{L}{m_l \bar{q}_l \gamma_5 q_l}$ 
and $Q(x)$ is the topological charge density, Eq. (\ref{EQN10.5})
can also be written in this way:
\be
2L \int d^4x \langle TQ(x)O_P^{(L)}(0) \rangle =
2Li \langle {\cal L}_{eff}^{(L)}(0) \rangle
- \int d^4x \langle TD_{(L)}(x)O_P^{(L)}(0) \rangle .
\label{EQN10.6}
\ee
If the $U(1)$ chiral symmetry is still broken above $T_{ch}$, i.e.,
$\langle {\cal L}_{eff}^{(L)}(0) \rangle \ne 0$ for $T > T_{ch}$
(while $\langle \bar{q} q \rangle = 0$ for $T > T_{ch}$), then this WI
implies the existence of a (pseudo--)Goldstone boson (in the large--$N_c$
limit!) coming from this breaking and interpolated by the hermitian
field $O_P$. Therefore, the $U(1)_A$ (pseudo--)NG boson (i.e.,
the $\eta'$) is an ``exotic'' $2L$--fermion state for $T > T_{ch}$!
The fact that $\langle {\cal L}_{eff}^{(L)}(0) \rangle $ is different from zero
above $T_{ch}$ in the chiral limit of zero quark masses, forces us to introduce
a new $2L$--fermion effective vertex with a coupling constant that we may
call ``$\gamma$''. This is shown schematically in Fig. 1.

\begin{figure}[htb]
\vskip 4.5truecm
\includegraphics{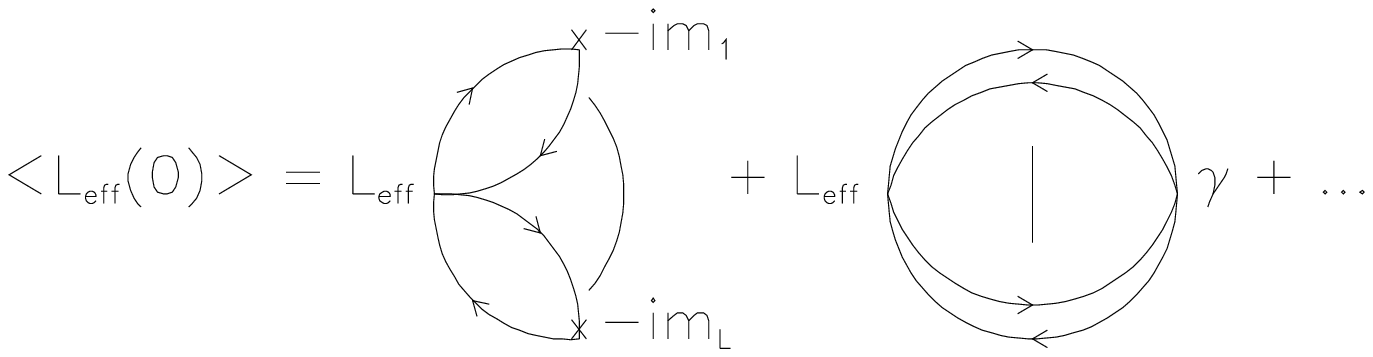}
\caption{The new $2L$--fermion effective vertex.}
\end{figure}

Now let us consider the implications of this picture on the saturation 
procedure of the other relevant QCD Ward Identities in the region of
temperatures between $T_{ch}$ and $T_\chi \le T_{U(1)}$, having assumed that
$T_{ch} < T_\chi \le T_{U(1)}$ \cite{EM1994b}. In other words, we suppose to
be in a region of temperatures $T>T_{ch}$ where the $SU(L) \otimes SU(L)$
chiral symmetry is restored (i.e., ${\langle \bar{q}_l q_l \rangle} = 0$,
in the chiral limit $\sup(m_l) \to 0$), the topological susceptibility of
the pure Yang--Mills (YM) theory is different from zero
(i.e., $\langle QQ \rangle |_{YM} \equiv i\lambda^2_{YM} \ne 0$;
here and in the following we make use of the notation:
$\langle AB \rangle \equiv \int {d^4 x \langle T A(x) B(0) \rangle}$)
and the $U(1)$ axial symmetry is still broken. Following Veneziano's model
for the resolution of the $U(1)$ problem \cite{Veneziano79}, 
we may get that $\langle QQ \rangle|_{YM} \equiv i\lambda^2_{YM}$
is different from zero by introducing an axial four--vector ghost of propagator
$-ig_{\mu \nu} / q^2$ coupled with strength $i\lambda_{YM}$ to the current 
$K_\mu$, whose divergence is the topological charge density $Q$
($Q \equiv \partial^\mu K_\mu$):
\be
\langle QQ \rangle |_{YM} =
\left\{ (i\lambda_{YM} )^2 (iq_\mu ){-ig^{\mu \nu } \over q^2 }
(-iq_\nu ) \right\} _{q \to 0} = i\lambda^2_{YM} .
\label{EQN10.1}
\ee
In order to explain, in a $1/N_c$ picture (see Ref. \cite{Witten79a}), how it
can be that $\langle QQ \rangle = 0$, in the chiral limit $\sup(m_l) \to 0$,
while $\langle QQ \rangle |_{YM}$ is of order ${\cal O}(1) \ne 0$ with respect
to the light quark masses, we have to consider (see Ref. \cite{Veneziano79})
that there are intermediate pseudo--scalar bound states which couple to the
ghost and which therefore must be resummed.
Since we are above the $SU(L) \otimes SU(L)$ chiral phase transition, it is
natural to suppose that there is just one $SU(L)$--singlet pseudo--scalar state
which plays such a role and which also breaks the $U(1)$ chiral symmetry.
Following Section 7, we take as an order parameter for the
$U(1)$ chiral symmetry alone the expectation value of the $2L$--fermion 
interaction term ${\cal L}_{eff}^{(L)}$ of the type of the one introduced
by 'tHooft in Ref. \cite{tHooft76}.
 
With the notation introduced before, the relevant WI's at $T \ne 0$ take
the form:
\be
\int d^4x \langle TD_l(x)Q(0) \rangle = 
-2\int d^4x \langle TQ(x)Q(0) \rangle ,
\label{EQN10.9}
\ee
\be
4 \int d^4x \langle TQ(x)Q(0) \rangle = 2i\delta_{lj}
\langle \Delta_l(0) \rangle
+\int d^4x \langle TD_l(x)D_j(0) \rangle .
\label{EQN10.10}
\ee
where: $\Delta_l (x) \equiv -2m_l \bar{q}_l q_l$ and
$D_l (x)\equiv 2i m_l \bar{q}_l \gamma_5 q_l$ (so that $D_{(L)} =
\sum_{l=1}^L{D_l}$).
We require their saturation at the leading order in a $1/N_c$ expansion.
To have a smooth limit when $N_c \to +\infty$, it is necessary not only to
take the QCD coupling constant ``$g$'' of the form $g=g_0/ \sqrt{N_c}$ with
$g_0$ of order ${\cal O}(1)$ with respect to $N_c$ (see Refs. \cite{tHooft74}
and \cite{Veneziano76}) but also to take the new coupling constant ``$\gamma$''
of the form:
\be
\gamma = { \gamma_0 \over N_c^{L-1} } ,
\label{EQN10.11}
\ee
with $\gamma_0$ still of order ${\cal O}(1)$ in $N_c$. One can easily be
convinced of this by analysing the $N_c$--dependence of the relevant Feynman
diagrams for Eqs. (\ref{EQN10.9}) and (\ref{EQN10.10}), considering also
all possible insertions of the new effective vertex ``$\gamma$''.
On the other hand, Eq. (\ref{EQN10.11}), with $\gamma_0$ of order ${\cal O}(1)$
in $N_c$, is guaranteed by a proper choice of the operator
${\cal L}^{(L)}_{eff}$, such that its expectation value $C^{(L)}_{U(1)} \equiv
\langle {\cal L}^{(L)}_{eff} \rangle$ comes out to be really independent of the
usual chiral condensate $\langle \bar{q} q \rangle$, as discussed at the end of
Section 7: in fact, this implies $C^{(L)}_{U(1)} = {\cal O}(N_c)$, from which
one immediately derives Eq. (\ref{EQN10.11}), after a comparison with Fig. 1.

\begin{figure}[htb]
\vskip 4.5truecm
\includegraphics{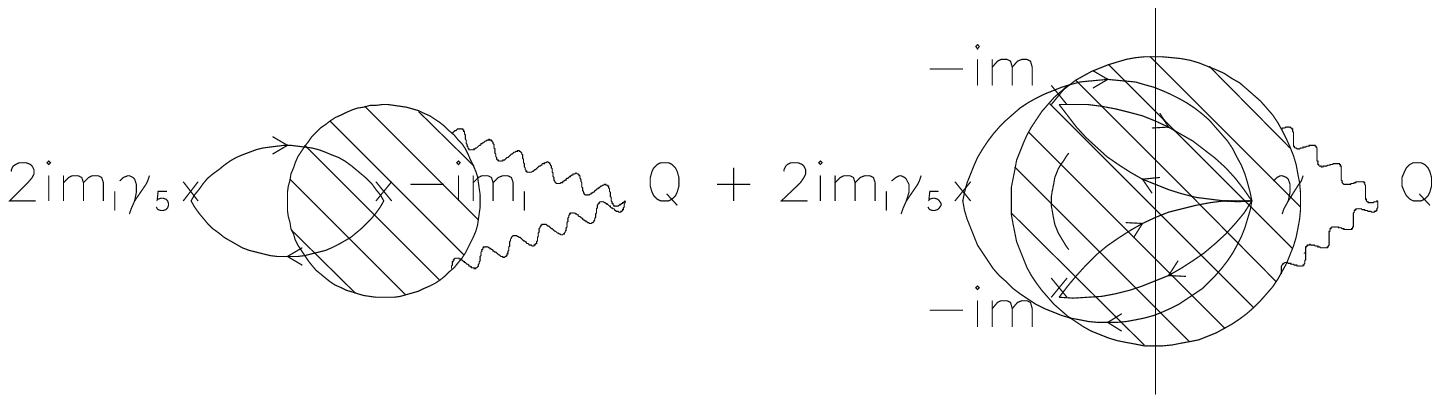}
\caption{The LHS of Eq. (\ref{EQN10.9}).}
\end{figure}

Let us consider the WI (\ref{EQN10.9}), which we briefly write as 
$\langle D_l Q \rangle = -2 \langle QQ \rangle$.
At the leading order in $1/N_c$ the left--hand side (LHS) term is 
schematically of the form given in Fig. 2.
Shaded blobs represent the sum of all diagrams of a given topology
(i.e., what we may call a ``{\it gluonic covering}'' of the basic 
diagrams).
The first piece is of order ${\cal O}(1)$ in $N_c$ and of order
${\cal O}(m_l^2)$ with respect to the light quark masses; the second one is of
order ${\cal O}(\gamma N_c^{L-1}) = {\cal O}(1)$ in $N_c$ and of order
${\cal O}(\prod_{k=1}^L{m_k})$ with respect to the quark masses.
The right--hand side (RHS) of Eq. (\ref{EQN10.9}) at the leading order in 
$1/N_c$ is represented in Fig. 3.

\begin{figure}[htb]
\vskip 4.5truecm
\includegraphics{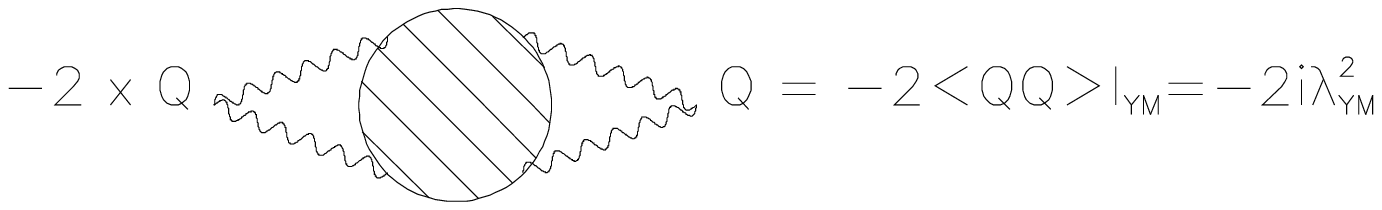}
\caption{The RHS of Eq. (\ref{EQN10.9}).}
\end{figure}

\noindent
It is of order ${\cal O}(1)$ in $N_c$ (i.e., $\lambda_{YM} = {\cal O}(1)$) and
of order ${\cal O}(1)$ with respect to the quark masses. How can one
match the (LHS) with the (RHS) of the WI (\ref{EQN10.9}) at the leading
order in $1/N_c$?
We observe that the second diagram in Fig. 2 contains, as an intermediate 
state, just the pseudoscalar $SU(L)$--singlet boson described by the field
$O_P^{(L)}(x)$. This is a new ``exotic'' particle which has essentially the 
same quantum numbers (spin, parity and so on) as the $\eta'$ at $T=0$: 
yet it has a completely different quark content, being a $2L$--fermion 
field of the form $\sim i[ {\det}(\bar{q}_{sL}q_{tR})
- {\det}(\bar{q}_{sR}q_{tL})]$.
We may obtain that the WI (\ref{EQN10.9}) is saturated by this bound state
$|S_X\rangle$, assuming that it couples to the four--vector ghost of $K_\mu$
with strength $if_s{1 \over \sqrt{\gamma N_c^L}}q^\mu$
($q^\mu$ being the four--momentum), where $f_s$ is of order ${\cal O}(1)$
in $N_c$. (The fact that $f_s$ is of order
${\cal O}(1)$ in $N_c$ will be verified below, as a consequence of
Eq. (\ref{EQN10.13}).) It also couples to the 
operator $D_l$ with strength $m_0^2 F_{D_l}$ (observe that $D_l$ is the
divergence of a four--current: $D_l=\partial^\mu (J^l_{5, \mu} - 2K_\mu)$,
where $J^l_{5, \mu} = \bar{q}_l \gamma_\mu \gamma_5 q_l$):
\be
\langle 0|D_l|S_X \rangle = m^2_0 F_{D_l} ,
\label{EQN10.12}
\ee
where $m_0$ is the mass of the state $|S_X\rangle$ at this OZI order of 
the expansion (i.e., before doing the resummation). We obtain that:
\ba
{\rm LHS}(\ref{EQN10.9}) &=& \left\{ (F_{D_l}m^2_{0}) {i \over q^2 - m^2_{0}}
\left( if_s{1 \over \sqrt{\gamma N_c^L}} q_\mu \right)
{-ig^{\mu \nu } \over q^2} i\lambda_{YM} (-iq_\nu ) \right\}_{q \to 0}
\nonumber \\
&=& -iF_{D_l} {f_s \over \sqrt{\gamma N_c^L}} \lambda_{YM} .
\label{EQN10.13}
\ea
$F_{D_l}$ is of order ${\cal O}(\sqrt{\gamma N_c^L}) = {\cal O}(\sqrt{N_c})$.
This can be seen considering the relevant Feynman diagrams, in the correlation
function $\langle D_l D_l \rangle$, which contain the state $|S_X\rangle$ as an
intermediate (propagating) state (the procedure is the same which is also used
in the usual ``{\it quark--anti-quark}'' meson case, and it is well described
in the first part of Ref. \cite{Witten79b}): these relevant diagrams contain
two $\gamma$--vertex insertions and, by virtue of Eq. (\ref{EQN10.11}), they
are of order ${\cal O}(N_c)$ with respect to $N_c$. Since, as we have seen
above, Eq. (\ref{EQN10.13}) is of order ${\cal O}(1)$ with respect to $N_c$,
we immediately derive that $f_s$ is of order ${\cal O}(1)$.

Comparing the expression (\ref{EQN10.13}) with the RHS of Eq. (\ref{EQN10.9}),
written in Fig. 3, we derive that:
\be
F_{D_l} = {2\lambda_{YM} \sqrt{\gamma N_c^L} \over f_s} .
\label{EQN10.14}
\ee
As it must be, if we want to saturate the WI (\ref{EQN10.9}), the coupling
constant $F_{D_l}$ has the same value ($F_D$) for every index $l=1,2,\ldots,L$.
From the structure of the second piece of Fig. 2 and from
Eq. (\ref{EQN10.12}), one immediately recognizes that the squared OZI mass
$m_0^2$ is proportional to the product of the $L$ light quark masses
(assuming, as usual, that $F_{D_l}$ is of order ${\cal O}(1)$ with respect
to them):
\be
m^2_{0} = O\left( \displaystyle\prod_{k=1}^L{m_k} \right) .
\label{EQN10.15}
\ee
All of this comes directly from the leading--order terms of the expansion. 
All quark loops, inserted in the leading non--OZI topology, 
can then be easily resummed as a geometric series and we get that:
\ba
\lefteqn{
\langle QQ \rangle^{F.T.} (q) \equiv \int d^4x e^{iqx} \langle TQ(x)Q(0)
\rangle = } \nonumber \\
& & =i\lambda^2_{YM} \left( 1+ {f^2_s \over \gamma N_c^L}
{1 \over q^2 - m^2_{0}} + \ldots \right) =
i\lambda^2_{YM} { q^2 -m^2_{0} \over q^2 - (m^2_{0} + 
{f^2_s \over \gamma N_c^L} ) } .
\label{EQN10.16}
\ea
Also, in the same way:
\ba
\lefteqn{
\langle D_l Q \rangle^{F.T.} (q) \equiv \int d^4x e^{iqx}\langle TD_l(x)Q(0)
\rangle = } \nonumber \\
& & =iF_{D_l}m^2_{0}{f_s \over \sqrt{\gamma N_c^L}}
\lambda_{YM} {1 \over q^2 - m^2_{0} }
\left( 1+ {f^2_s \over \gamma N_c^L}{1 \over q^2 - m^2_{0}} + \ldots \right) =
\nonumber \\
& & =iF_{D_l}m^2_{0}{f_s \over \sqrt{\gamma N_c^L}}
\lambda_{YM} {1 \over q^2 - (m^2_{0} + {f^2_s \over \gamma N_c^L} ) } .
\label{EQN10.17}
\ea
The right--hand sides of Eqs. (\ref{EQN10.16}) and (\ref{EQN10.17}) have
a pole corresponding to the physical mass of the pseudoscalar singlet
boson $|S_X\rangle$:
\be
m^2_S = m^2_{0} + {f^2_s \over \gamma N_c^L} 
= m^2_{0} + {4 \lambda^2_{YM} \over F^2_D} .
\label{EQN10.18}
\ee
(In the last passage we have used Eq. (\ref{EQN10.14}).)
So the boson $|S_X\rangle$ acquires a squared ``topological'' mass $f_s^2/
\gamma N_c^L = 4 \lambda^2_{YM} / F^2_D$, which is different from zero also 
in the chiral limit of zero quark 
masses (when, on the contrary, the OZI mass $m_0$ vanishes).
Moreover this squared mass is of order ${\cal O}(1/N_c)$ in the number of
colours, just like the anomaly term and the $\eta'$ squared 
mass in the Witten--Veneziano
model \cite{Witten79a,Veneziano79}: in this sense we may say that this
new ``{\it exotic}'' state is a ``{\it light}'' state, at large $N_c$.
This result will also be obtained in Section 14 in an effective Lagrangian
model. It is also easy to verify, using
Eq. (\ref{EQN10.14}), that the expressions (\ref{EQN10.16}) and
(\ref{EQN10.17}), in the limit $q \to 0$, satisfy the WI (\ref{EQN10.9})
(that is: $\langle D_l Q \rangle = -2 \langle QQ \rangle$);
they are both of order ${\cal O}(\gamma N_c^L) = {\cal O}(N_c)$ with respect
to $N_c$ and of order ${\cal O}(\prod_{k=1}^L{m_k})$ with respect
to the light quark masses.  In fact, using Eqs. (\ref{EQN10.14}) and
(\ref{EQN10.15}), we find:
\ba
\lefteqn{
\langle QQ \rangle \equiv \langle QQ \rangle^{F.T.} (q \to 0) =
i\lambda^2_{YM} {m^2_{0} \over m^2_{0} + {f^2_s \over \gamma N_c^L}} }
\nonumber \\
& & \mathop{\longrightarrow}_{\sup(m_l) \to 0}
i\lambda^2_{YM} {\gamma N_c^L \over f^2_s }m^2_{0} =
i{F^2_D \over 4}m^2_0 =
{\cal O} \left( \displaystyle\prod_{k=1}^L{m_k} \right) .
\label{EQN10.19}
\ea
Also this result is in total agreement with what we will find in Section 14
using an effective Lagrangian model.
And what about the chiral condensate $\langle \bar{q}_l q_l \rangle$?
At the leading order in our expansion in $N_c$ and in the light 
quark masses $m_l$ ($l=1,2,\ldots ,L$) it receives two different contributions 
of order ${\cal O}(N_c)$: one from a mass insertion and another from a
$\gamma$--vertex insertion (see Fig. 4).

\begin{figure}[htb]
\vskip 4.5truecm
\includegraphics{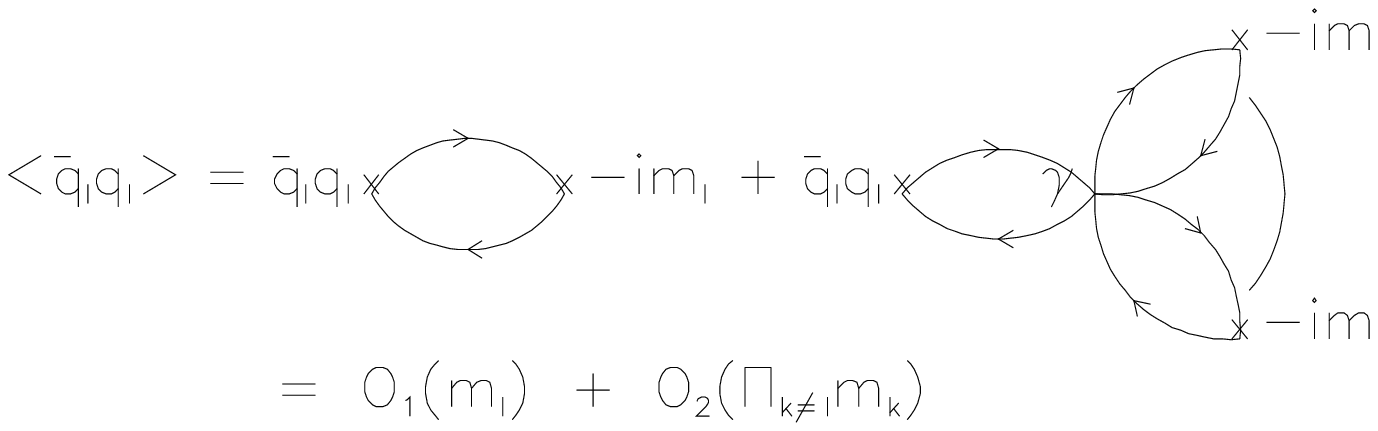}
\caption{The chiral condensate above $T_{ch}$.}
\end{figure}

\noindent
Also this last behavior of the chiral condensate with respect to the
light--quark masses will be found in Section 14. (In the Appendix we also 
derive the same quark--masses dependence for the topological 
susceptibility and the chiral condensate using a simple instantonic
model.) As a curiosity, the fact that the chiral condensate $\langle \bar{q}_l
q_l \rangle$ does not vanish as the mass $m_l$ goes to zero (since there
remains the piece ${\cal O}_2$ proportional to the product of the remaining
$L-1$ quark masses) could appear to be rather unnatural. Yet it is not so
strange, if one considers that the expectation value
$\langle {\cal L}_{eff}^{(L)}(0) \rangle$ is an order parameter 
also of the $U(1)_A^{(l)}$ chiral symmetry for the flavour ``$l$'' (i.e.,
the group of transformations $q_l \to e^{-i \alpha_l \gamma_5}q_l$,
which rotates only the flavour $q_l$). Therefore, if
$\langle {\cal L}_{eff}^{(L)}(0) \rangle \ne 0$, as in our case, the
$U(1)_A^{(l)}$ symmetry is broken even when the mass $m_l$ of the flavour
$q_l$ goes to zero. Finally, we can immediately verify that also the second
relevant WI (\ref{EQN10.10}) is saturated:
\be
4 \langle QQ \rangle = 2i\delta_{lj} \langle \Delta_l(0) \rangle +
\langle D_l D_j \rangle .
\label{EQN10.20}
\ee
Let us take $l=j$ in the last equation. Then it is easy to recognize that 
the ${\cal O}_1(m_l)$ term in Fig. 4, that is the part of order
${\cal O}(N_c)$ in $N_c$ and of order ${\cal O}(m_l^2)$ in the quark masses for
$\langle \Delta_l(0) \rangle$, cancels exactly with the analogous part of
$\langle D_l D_l \rangle$:
\be
2i\langle \Delta_l(0) \rangle |_{N_c, m^2_l} + \langle D_l D_l \rangle
|_{N_c, m^2_l} =0 .
\label{EQN10.21}
\ee
The part of $\langle \Delta_l(0) \rangle$ of order ${\cal O}(\gamma N_c^L) =
{\cal O}(N_c)$ in $N_c$ and of order
${\cal O}(\prod_{k=1}^L{m_k})$ in the quark masses,
coming from the ${\cal O}_2(\prod_{k \ne l}{m_k})$ term in
Fig. 4, is forced to be directly proportional to the ``full'' topological
susceptibility (since $\langle D_l D_l \rangle |_{\gamma N_c^L,
\prod_{k=1}^L{m_k}} = 0$):
\be
2i \langle \Delta_l(0) \rangle \mid_{\gamma N_c^L, \prod_{k=1}^L{m_k}} =
-4im_l{\cal O}_2 \left( \displaystyle\prod_{k \ne l}{m_k} \right) =
4 \langle QQ \rangle .
\label{EQN10.22}
\ee
That is: $\langle QQ \rangle = -im_l {\cal O}_2(\prod_{k \ne l}{m_k})$. 
Also this relation will be found in Section 14, as a consequence of an
effective Lagrangian model.
For $l \ne j$ the result is the same, since one can easily verify that:
\be
\forall l \ne j : \ 
2i \langle \Delta_l(0) \rangle \mid_{\gamma N_c^L, \prod_{k=1}^L{m_k}} =
\langle D_l D_j \rangle |_{\gamma N_c^L, \prod_{k=1}^L{m_k}} ,
\label{EQN10.23}
\ee
in agreement with the fact that both of them must be equal to
$4\langle QQ \rangle$.

\newsection{The new chiral effective Lagrangian}

\noindent
We will see in this Section how the proposed scenario, in which the $U(1)$
axial symmetry is still broken above the chiral transition, can be
consistently reproduced using an effective--Lagrangian
model \cite{EM1994a,EM1994b,EM1994c}.

It is well known that the low--energy dynamics of the pseudoscalar mesons, 
including the effects due to the anomaly and expanding to the first order 
in the light quark masses, can be described, in the large--$N_c$ limit, 
by this kind of effective Lagrangian \cite{DiVecchia-Veneziano80,Witten80,
Rosenzweig-et-al.80,Nath-Arnowitt81,Leutwyler91}:
\ba
\lefteqn{
{\cal L}(U,U^\dagger ,Q) = {\cal L}_0(U,U^\dagger )
+ {B_m \over 2 \sqrt{2}} \Tr[MU + M^\dagger U^\dagger ] } \nonumber \\
& & + {1 \over 2}iQ(x) \Tr[\log(U) -\log(U^\dagger )] + {1 \over 2A}Q^2(x) ,
\label{EQN3.1}
\ea
where ${\cal L}_0$ describes a kind of linear $\sigma$--model:
\be
{\cal L}_0(U,U^\dagger )={1 \over 2}\Tr(\partial_\mu U\partial^\mu U^\dagger )
-{1 \over 4}\lambda_{\pi}^2 \Tr[(U^\dagger U-\rho_\pi \cdot {\bf I})^2].
\label{EQN3.2}
\ee
{\bf I} is the identity matrix.
A few words about the notation used in Eq. (\ref{EQN3.1}).
$M$ represents the quark mass matrix, which enters in the QCD Lagrangian
as $\delta {\cal L}^{(mass)}_{QCD} = -\bar{q}_R M q_L + {\rm h.c.}$, that is:
\be
M = \pmatrix{ m_1& & & \cr
& m_2& & \cr
& & \ddots& \cr
& & & m_L \cr}
.
\label{EQN3.3}
\ee
$Q(x)={g^2 \over 64\pi^2}\varepsilon^{\mu\nu\rho\sigma} F^a_{\mu\nu}(x)
F^a_{\rho\sigma}(x)$ 
is the topological charge density and $A$ is a real positive 
coefficient, which is essentially the topological susceptibility in the 
pure Yang--Mills (YM) theory (i.e., in the absence of quarks):
\be
\int d^4x \langle TQ(x)Q(0) \rangle |_{YM}=iA.
\label{EQN3.4}
\ee
The parameter $\rho_{\pi}$, which is a function of the temperature $T$, 
controls the behaviour of the theory across the chiral transition, being
$\rho_{\pi}(T < T_{ch})={1 \over 2}F_{\pi}^2 > 0$ and
$\rho_{\pi}(T > T_{ch})=-{1 \over 2}B_{\pi}^2 < 0$ 
($\rho_{\pi}(T=T_{ch})=0$). The mesonic field $U_{ij}$ is a $L \times L$ 
complex matrix, which, in terms of the quark fields $q_i$ ($i=1,2,\ldots ,L$)
is essentially:
\be
U_{ij} \sim \bar{q}_j \left( {1+\gamma_5 \over 2} \right) q_i
= \bar{q}_{jR} q_{iL} ,
\label{EQN3.5}
\ee
up to a multiplicative constant;
as usual, $q_L \equiv {1 \over 2}(1 + \gamma_5)q$ and
$q_R \equiv {1 \over 2}(1 - \gamma_5)q$,
with $\gamma_5 \equiv -i \gamma^0 \gamma^1 \gamma^2 \gamma^3$,
denote the {\it left--handed} and the {\it right--handed} quark fields.
Since under chiral $U(L) \otimes U(L)$ transformations the quark fields 
transform as follows:
\be
q_L \to V_L q_L ~~~,~~~q_R \to V_R q_R,
\label{EQN3.6}
\ee
where $V_L$ and $V_R$ are arbitrary $L \times L$ unitary matrices, 
we immediately derive the transformation property of
$U$ under $U(L) \otimes U(L)$:
\be
U \rightarrow V_L U V_R^\dagger .
\label{EQN3.7}
\ee
Let us also observe that ${\cal L}_0(U,U^{\dagger})$ is invariant under
$U(L) \otimes U(L)$, while the Lagrangian ${\cal L}(U,U^{\dagger} ,Q)$,
in the chiral limit $\sup(m_i) \to 0$ ($i=1,2,\ldots ,L$), is invariant under
$SU(L) \otimes SU(L) \otimes U(1)_V$ (``$V$'' stands for {\it ``vectorial''})
and it has the same transformation property of the massless QCD Lagrangian
${\cal L}^{(0)}_{QCD}$ under a $U(1)$ chiral transformation:
\be
U(1)_A:~~~ {\cal L}^{(0)}_{QCD}(x) \to {\cal L}^{(0)}_{QCD}(x)
+ 2L \alpha Q(x),
\label{EQN3.8}
\ee
where $\alpha$ is the parameter of the $U(1)$ chiral transformation 
($U(1)_A$: $q \to e^{-i\alpha\gamma_5}q$, i.e., $q_L \to e^{-i\alpha}q_L$
and $q_R \to e^{i\alpha}q_R$).
The Lagrangian (\ref{EQN3.1}) considers, as a chiral order parameter, only
the usual chiral condensate $\langle \bar{q} q \rangle$: where it is
different from zero, both $SU(L) \otimes SU(L)$ and $U(1)_A$ are broken. 
If we make the assumption that the $U(1)$ chiral symmetry is restored at a
temperature $T_{U(1)}$ greater than $T_{ch}$, we need another order parameter
for the $U(1)$ chiral symmetry, the form of which has been discussed in the
previous Sections. We must now define a field variable $X$,
associated with this new condensate, to be inserted in the chiral Lagrangian.
The translation from the fundamental quark fields to the
effective--Lagrangian meson fields is done as follows. The operators
$i \bar{q} \gamma_5 q$ and $\bar{q} q$ entering in the WI (\ref{eqn7})
are essentially equal to (up to a multiplicative constant)
$i(\Tr U - \Tr U^\dagger)$ and $\Tr U + \Tr U^\dagger$ respectively.
Similarly, the operators
${\cal L}_{eff} \sim {\det}(\bar{q}_{sR}q_{tL}) + {\det}(\bar{q}_{sL}q_{tR})$
and $O_P \sim i[ {\det}(\bar{q}_{sR}q_{tL}) - {\det}(\bar{q}_{sL}q_{tR}) ]$
entering in the WI (\ref{EQN10.5}) can be put equal to (up to a multiplicative
constant) $X + X^\dagger$ and $i(X - X^\dagger)$ respectively, where
\be
X \sim \displaystyle{\det_{st}} \left[ \bar{q}_s \left( {1+\gamma_5 \over 2}
\right) q_t \right]
= \displaystyle{\det_{st}} \left( \bar{q}_{sR} q_{tL} \right)
\label{EQN3.9}
\ee
is the new field variable (up to a multiplicative constant),
related to the new $U(1)$ chiral condensate, which must be inserted
in the chiral effective Lagrangian.
Clearly $X$ has the following transformation properties under
$U(L) \otimes U(L)$:
\be
U(L) \otimes U(L):~~~ X \to \det(V_L)\det(V_R)^* X.
\label{EQN3.10}
\ee
That is, $X$ is invariant under $SU(L) \otimes SU(L) \otimes U(1)_V$, 
while, under $U(1)_A$, $X \to e^{-i 2L \alpha} X$.

The new Lagrangian, written in terms of the new fields $U_{ij}$,
$X$ and $Q$, must have the following properties:
\begin{itemize}
\item{} it must be invariant under $U(L) \otimes U(L)$, if one neglects 
the quark masses and the effects due to the anomaly;
\item{} it must transform in the same way as ${\cal L}^{(0)}_{QCD}$ in
(\ref{EQN3.8}) under the group $U(1)_A$ (always neglecting the quark masses).
\end{itemize}
As usual we will consider the effects of the quark masses by making an 
expansion in powers of $M$ and taking only the linear term. Since the 
effective Lagrangian has to be invariant under a 
transformation of $U$ as in (\ref{EQN3.7}) and of $X$ as in (\ref{EQN3.10})
together with a transformation of $M$ of the form
(we recall that $M$ represents the quark mass matrix, which enters in the
QCD Lagrangian as $\delta {\cal L}^{(mass)}_{QCD} = -\bar{q}_R M q_L +
{\rm h.c.}$):
\be
M \to V_R M V_L^\dagger ,
\label{EQN3.11}
\ee
we see that the most simple term linear in $M$ is always of the 
following form:
\be
\delta {\cal L}_{mass} = {B_m \over 2 \sqrt{2}}\Tr[MU +M^\dagger U^\dagger].
\label{EQN3.12}
\ee
At this point it is rather simple to see that the most simple effective 
Lagrangian, constructed with the fields $U$, $X$ and $Q$ and satisfying all 
the properties listed above, is \cite{EM1994a}:
\ba
\lefteqn{
{\cal L}(U,U^\dagger ,X,X^\dagger ,Q) =
{1 \over 2}\Tr(\partial_\mu U\partial^\mu U^\dagger )
+ {1 \over 2}\partial_\mu X\partial^\mu X^\dagger + } \nonumber \\
& & -V(U,U^\dagger ,X,X^\dagger ) +{1 \over 2}iQ(x)\omega_1 \Tr[\log(U) - 
\log(U^\dagger )]+ \nonumber \\
& & +{1 \over 2}iQ(x)(1-\omega_1)[\log(X)-\log(X^\dagger )]+{1 \over 2A}Q^2(x) ,
\label{EQN3.13}
\ea
where the potential term $V(U,U^{\dagger},X,X^{\dagger})$ has the form:
\ba
\lefteqn{
V(U,U^\dagger ,X,X^\dagger )={1 \over 4}\lambda_{\pi}^2 \Tr[(U^\dagger U
-\rho_\pi \cdot {\bf I})^2] +
{1 \over 4}\lambda_X^2 (X^\dagger X-\rho_X )^2 } \nonumber \\
& & -{B_m \over 2\sqrt{2}}\Tr[MU+M^\dagger U^\dagger ]
-{c_1 \over 2\sqrt{2}}[\det(U)X^\dagger + \det(U^\dagger )X] .
\label{EQN3.14}
\ea
All the parameters appearing in the Lagrangian must be considered as 
functions of the physical temperature $T$. In particular, the parameters 
$\rho_{\pi}$ and $\rho_X$ are responsible for the behaviour of the theory 
respectively across the $SU(L) \otimes SU(L)$ 
and the $U(1)$ chiral phase transitions, according to the following table:
$$
\vbox{\tabskip=0pt \offinterlineskip
\halign to320pt{\strut#
& \vrule#\tabskip=1em plus2em& \hfil#&\vrule#
& \hfil#&\vrule#
& \hfil#&\vrule#
& \hfil#&\vrule#
\tabskip=0pt\cr\noalign{\hrule}
&& \omit\hidewidth { } \hidewidth
&& \omit\hidewidth $T<T_{ch}$ \hidewidth
&& \omit\hidewidth $T_{ch}<T<T_{U(1)}$ \hidewidth
&& \omit\hidewidth $T>T_{U(1)}$ \hidewidth& \cr
\noalign{\hrule}
&& { } & & { } & & { } & & { } &\cr
&& $\rho_\pi$ & & ${1\over 2}F_\pi^2>0$ & & $-{1\over 2}B_\pi^2<0$ & 
& $-{1\over 2}B_\pi^2<0$ &\cr 
&& { } & & { } & & { } & & { } &
\cr\noalign{\hrule}
&& { } & & { } & & { } & & { } &\cr
&& $\rho_X$ & & ${1\over 2}F_X^2>0$ & & ${1\over 2}F_X^2>0$ & 
& $-{1\over 2}B_X^2<0$ &\cr 
&& { } & & { } & & { } & & { } &
\cr\noalign{\hrule}\noalign{\smallskip}& \multispan7 [Tab.1]
\hfil\cr}}
$$
(That is: $\rho_{\pi}(T_{ch})=0$ and $\rho_X(T_{U(1)})=0$.)
The $U(1)$ chiral symmetry remains broken also in the region of 
temperatures $T_{ch} < T < T_{U(1)}$, where on the contrary the 
$SU(L) \otimes SU(L)$ 
chiral symmetry is restored. The $U(1)$ chiral symmetry is restored 
above $T_{U(1)}$. Following the Introduction, we also assume that the 
topological susceptibility of the pure YM theory, $A(T)$, drops to zero at 
a temperature $T_{\chi}$ greater than $T_{ch}$. We will see later that, for 
the consistency of the model, it must be $T_{\chi} \le T_{U(1)}$.
In the following it will be sometimes useful to integrate out the field 
variable $Q(x)$ in the Lagrangian (\ref{EQN3.13}), obtaining:
\ba
\lefteqn{
{\cal L}(U,U^\dagger ,X,X^\dagger ) =
{1 \over 2}\Tr(\partial_\mu U\partial^\mu U^\dagger )
+ {1 \over 2}\partial_\mu X\partial^\mu X^\dagger + }
\nonumber \\
& & -V(U,U^\dagger ,X,X^\dagger ) +{1 \over 8}A\{\omega_1
\Tr[\log(U) - log(U^\dagger )]+ \nonumber \\
& & +(1-\omega_1)[\log(X)-\log(X^\dagger )]\}^2 .
\label{EQN3.15}
\ea
In order to avoid a singular behaviour of the anomaly term in (\ref{EQN3.15})
just above $T_{ch}$, where the ``{\it vacuum}'' expectation value of $U$ tends 
to zero in the chiral limit $\sup(m_i) \to 0$ and where $A(T) > 0$, we will 
assume that:
\be
\omega_1(T \ge T_{ch}) = 0.
\label{EQN3.16}
\ee

\newsection{Mass spectrum of the theory for $T_{ch}<T<T_{U(1)}$}

\noindent
Let us now consider the Lagrangian (\ref{EQN3.15}), where the field variable
$Q(x)$ has been integrated out; we study the mass spectrum of the theory in
the region of temperatures $T_{ch} < T < T_{U(1)}$. As it has been stressed in
Ref. \cite{DiVecchia-Veneziano80}, the linear $\sigma$--type model contains
redundant scalar fields, 
which can be eliminated by taking the limit $\lambda_{\pi}^2 \to +\infty$ 
and $\lambda_X^2 \to +\infty$ in Eq. (\ref{EQN3.14}): so we will expand our
results not only in powers of the light quark masses $m_i$ ($i=1,2,\ldots ,L$),
but also in powers of $1/\lambda_{\pi}^2$ and $1/\lambda_X^2$. In the region of
temperatures that we are considering, we have, according to Table 1,
$\rho_\pi =-{1 \over 2}B_\pi^2 < 0$ and $\rho_X = {1 \over 2} F_X^2 > 0$.
So, because of the form of the potential term (\ref{EQN3.14}), in this region
the $SU(L) \otimes SU(L)$ chiral symmetry is restored, being
$\langle U^{\dagger}U \rangle = 0$, in the chiral limit $\sup(m_i) \to 0$,
while the $U(1)$ chiral symmetry is still broken, being
$\langle X^{\dagger}X \rangle ={1 \over 2}F_X^2$.
Moreover, from Eq. (\ref{EQN3.16}) we have that $\omega_1(T>T_{ch})=0$.
If we parametrize the complex fields $U$ and $X$ by separating 
their {\it real} and {\it imaginary} parts in the following way:
\be
U_{ij}=a_{ij}+ib_{ij}~~~;~~~X=\theta_1 +i\theta_2,
\label{EQN4.1}
\ee
we immediately find that the minimum of the potential is obtained in the 
``point'' $P$ where:
\be
\theta_1|_P = \sqrt{{1 \over 2}}F_X~~;~~\theta_2|_P=0~~;~~
U|_P= {2B_m \over \sqrt{2}\lambda_\pi^2 B_\pi^2} M^\dagger.
\label{EQN4.2}
\ee
We first neglect the anomaly term in (\ref{EQN3.15}), that is the term
proportional to the topological susceptibility $A$ of the pure YM theory, and
then we study the mass spectrum of the theory, by diagonalizing the matrix of
the second derivatives in the ``point'' $P$ of the potential
$V(U,U^{\dagger},X,X^ {\dagger})$ with respect to the various fields
$a_{ij},~b_{ij},~\theta_1$ and $\theta_2$.
We then find that, in this limit, there is a field $S_X$ 
(which is essentially $\theta_2-\theta_2|_P=\theta_2$) which has a squared 
mass $m^2_0$ proportional to the product of the light quark masses:
\be
m^2_0 = {c_1 \over F_X} \left( {2B_m \over \sqrt{2}\lambda_\pi^2 B_\pi^2}
\right) ^L \det(M).
\label{EQN4.3}
\ee
All the other fields have squared masses equal to 
$\lambda_X^2 F_X^2$ or ${1\over 
2}\lambda_{\pi}^2 F_{\pi}^2$, so that, in the limit 
$\lambda_{\pi}^2 \to +\infty$ and $\lambda_X^2 \to +\infty$, they are 
redundant fields.
If we now consider the full potential $\tilde V(U,U^{\dagger},X,X^{\dagger})$ 
which appears in Eq. (\ref{EQN3.15}) as the sum of
$V(U,U^{\dagger},X,X^{\dagger})$ and of the anomaly term:
\be
\tilde{V}(U,U^\dagger ,X,X^\dagger ) = V(U,U^\dagger ,X,X^\dagger )
-{1 \over 8}A[\log(X)-\log(X^\dagger )]^2,
\label{EQN4.4}
\ee
we obtain that the field $S_X$ acquires a ``topological'' squared mass 
equal to ${2A \over F_X^2}$. In conclusion, the squared mass of the field 
$S_X$ is given by:
\be
m^2_{S_X} = m^2_0 + {2A \over F_X^2} \mathop{\longrightarrow}_{\sup(m_i)\to 
0}{2A \over F_X^2}.
\label{EQN4.5}
\ee
The physical interpretation of this new particle described by the field $S_X$
is rather obvious. It is nothing but the {\it would--be} Goldstone particle
coming from the breaking of the $U(1)$ chiral symmetry: in fact, neglecting 
the anomaly, it has zero mass in the chiral limit of zero quark masses.
Yet it acquires a ``topological'' squared mass proportional to the topological 
susceptibility $A(T)$ of the pure YM theory, just as the $\eta'$ in the 
Witten--Veneziano model \cite{Witten79a,Veneziano79}. 
And actually it has the same quantum 
numbers (spin, parity, etc.) as the $\eta'$, even if it is a sort of
``exotic'' matter field of the form
$\sim i[ {\det}(\bar{q}_{sL}q_{tR}) - {\det}(\bar{q}_{sR}q_{tL})]$.
Its existence could be proved perhaps in the near
future by heavy--ions experiments. 

Finally we also observe that, in order to avoid a singular behaviour of 
$m^2_{S_X}$ as the temperature $T$ approaches the $U(1)$ chiral transition 
temperature $T_{U(1)}$ (where $F_X$ vanishes), we must require that the 
``topological'' temperature $T_\chi$, at which $A(T)$ drops to zero, is not 
greater than $T_{U(1)}$; that is:
\be
T_{ch} < T_\chi \le T_{U(1)}.
\label{EQN4.6}
\ee

\newsection{Mass spectrum of the theory for $T<T_{ch}$}

\noindent
In the previous Section we have studied the mass spectrum of the theory for
$T_{ch} < T < T_{U(1)}$. It is obviously interesting to see also what 
happens for $T<T_{ch}$, where both the $SU(L) \otimes SU(L)$ 
and the $U(1)$ chiral symmetry are broken.
In fact, owing to the form of the potential (\ref{EQN3.14}), we 
have, at the leading order in $m_i$, $1/\lambda_{\pi}^2$ and 
$1/\lambda_X^2$:
\be
\langle U^\dagger U \rangle = {1 \over 2}F_\pi^2 \cdot {\bf I} ~~~ ; ~~~ 
\langle X^\dagger X \rangle = {1 \over 2}F_X^2 .
\label{EQN5.1}
\ee
As we have already explained in the previous Section, we can eliminate the 
redundant scalar fields of the linear $\sigma$--type model by taking the 
limit $\lambda_{\pi}^2 \to +\infty$ and $\lambda_X^2 \to +\infty$ in 
Eq. (\ref{EQN3.14}). In this limit the potential term gives the following
constraints:
\be
U^\dagger U = {1 \over 2}F_\pi^2 \cdot {\bf I} ~~~ ; ~~~ 
X^\dagger X = {1 \over 2}F_X^2 .
\label{EQN5.2}
\ee
So we can take the field $U$ of the form:
\be
U= \sqrt{1 \over 2}F_\pi \exp \left( {i\sqrt{2} \over F_\pi} \Phi \right) ,
\label{EQN5.3}
\ee
with:
\be
\Phi_{ij} = \displaystyle\sum_{\alpha \ne \beta}
\tilde \pi^{\alpha \beta} \tilde \tau^{\alpha \beta}_{ij}
+ \tilde \pi^i \delta_{ij} ,
\label{EQN5.4}
\ee
where the matrices $\tilde \tau_{ij}^{\alpha \beta}$ are the $L(L-1)$ 
non--diagonal generators of $U(L)$ [or, alternatively:
$\Phi = \sum_{a = 1}^{L^2-1} \pi_a \tau_a + {S_\pi \over \sqrt{L}}{\bf I}$,
where $\tau_a$ ($a=1,\ldots,L^2-1$) are the 
generators of the algebra of $SU(L)$ in the fundamental representation,
with normalization $\Tr(\tau_a) = 0$, $\Tr(\tau_a \tau_b) = \delta_{ab}$,
and $S_\pi$ is the $SU(L)$--singlet field].
And also:
\be
X= \sqrt{1 \over 2}F_X \exp \left( {i\sqrt{2} \over F_X} S_X \right) .
\label{EQN5.5}
\ee
Substituting Eqs. (\ref{EQN5.3}) and (\ref{EQN5.5}) into Eq. (\ref{EQN3.15})
(where the field variable $Q(x)$ has been already integrated) and taking only
the quadratic part of the Lagrangian, we obtain the following expression:
\ba
\lefteqn{
{\cal L}_2 ={1 \over 2}\displaystyle\sum_{i=1}^L{\partial_\mu \tilde\pi_i
\partial^\mu \tilde\pi_i} 
+{1 \over 2}\displaystyle\sum_{\alpha \ne \beta}{\partial_\mu \tilde\pi^
{\alpha \beta} \partial^\mu \tilde\pi^{\alpha \beta}} 
+{1 \over 2}\partial_\mu S_X \partial^\mu S_X + } 
\nonumber \\
& &-{1 \over 2}\displaystyle\sum_{i=1}^L {\mu_i^2 \tilde\pi_i^2}
-{1 \over 2}\displaystyle\sum_{\alpha \ne \beta}{{\mu_\alpha^2 +\mu_\beta^2 
\over 2} \tilde\pi^{\alpha \beta} \tilde\pi^{\alpha \beta}} +
\nonumber \\
& &-{c_1 \over 2\sqrt{2}}\left( {F_X \over \sqrt{2}}\right)
\left( {F_\pi \over \sqrt{2}}\right)^L
\left( {\sqrt{2} \over F_\pi}\displaystyle\sum_{i=1}^L{\tilde\pi_i}
-{\sqrt{2} \over F_X}S_X \right)^2 +
\nonumber \\
& &-A\left( {\omega_1 \over F_\pi}\displaystyle\sum_{i=1}^L{\tilde\pi_i}
+ {1-\omega_1 \over F_X}S_X \right)^2 ,
\label{EQN5.6}
\ea
where:
\be
\mu_i^2 \equiv {B_m \over F_\pi}m_i .
\label{EQN5.7}
\ee
If we also put, for simplicity:
\be
c \equiv {c_1 \over \sqrt{2}}\left( {F_X \over \sqrt{2}}\right) 
\left( {F_\pi \over \sqrt{2}} \right)^L ,
\label{EQN5.8}
\ee
we have the following squared mass matrix for the system of fields
($S_X$, $\tilde \pi_1$, $\tilde \pi_2$ ,$\ldots$ , $\tilde \pi_L$):
\be
{\bf A} =\pmatrix{
{2A(1-\omega_1)^2+2c \over F_X^2} & {2A\omega_1(1-\omega_1)-2c \over F_\pi 
F_X} & \ldots & {2A\omega_1(1-\omega_1)-2c \over F_\pi F_X} \cr
{2A\omega_1(1-\omega_1)-2c \over F_\pi F_X} & 
\mu^2_1 + {2A\omega_1^2 +2c \over F_\pi^2} & \ldots & 
{2A\omega_1^2 +2c \over F_\pi^2} \cr
\vdots & \vdots & \ddots & \vdots \cr
{2A\omega_1(1-\omega_1)-2c \over F_\pi F_X} &
{2A\omega_1^2 +2c \over F_\pi^2} & \ldots &
\mu_L^2 + {2A\omega_1^2 +2c \over F_\pi^2} \cr} .
\label{EQN5.9}
\ee
It is not hard to find out the $L+1$ eigenvalues of this matrix. In the 
chiral limit, in which the masses $m_i$ of the $L$ light quarks are put 
equal to zero (so that $\mu_i^2=0$; $i=1,2,\ldots ,L$) the matrix {\bf A} has 
$L-1$ null eigenvalues ($\lambda_0=\lambda_1=\ldots =\lambda_{L-2}=0$) and two 
other eigenvalues given by:
\be
\left. \matrix{\lambda_{L-1} \cr \lambda_L \cr } \right\} = 
{ Z_L \mp \sqrt{Z_L^2 -
4Q_L} \over 2},
\label{EQN5.10}
\ee
where $Z_L$ and $Q_L$ are so defined:
\ba
Z_L &\equiv& {2A[F_\pi^2(1-\omega_1)^2+LF_X^2\omega_1^2]+2c(F_\pi^2+LF_X^2)
\over F_\pi^2 F_X^2} , \nonumber \\
Q_L &\equiv& {4LAc \over F_\pi^2 F_X^2} .
\label{EQN5.11}
\ea
In other words, we have $L-1$ zero--mass states from the ensemble
$(S_X, \tilde \pi_1, \tilde \pi_2, \ldots ,\tilde \pi_L)$ which, together
with the $L(L-1)$ zero--mass states $\tilde \pi^{\alpha \beta}$,
constitute the $L^2-1$ Goldstone bosons coming from the breaking of the 
$SU(L) \otimes SU(L)$ chiral symmetry down to $SU(L)_V$. 
Then we have two non-zero--mass states: let us discuss them.
In Section 14 (after having derived some useful expressions for the chiral 
condensate) we will see that, in the limit of large number of 
colours $N_c$, we have the following $N_c$--dependences:
\be
F_\pi = {\cal O}(N_c^{1/2});~~ F_X = {\cal O}(N_c^{1/2});~~ A = {\cal O}(1);~~
c = {\cal O}(N_c).
\label{EQN5.12}
\ee
The two eigenstates $\phi_1$ and $\phi_2$ corresponding to $\lambda_{L-1}$
and $\lambda_L$ are, in the $N_c \to \infty$ limit:
\ba
\phi_1 &=& {1 \over \sqrt{F_\pi^2 + LF_X^2}}(\sqrt{L}F_X S_X + F_\pi S_\pi), 
\nonumber \\
\phi_2 &=& {1 \over \sqrt{F_\pi^2 + LF_X^2}}(-F_\pi S_X + \sqrt{L}F_X S_\pi), 
\label{EQN5.14}
\ea
where $S_\pi$ is the usual ``quark--anti-quark'' $SU(L)$--singlet field
(in terms of the fields $\tilde \pi_i$ it is given by: $S_\pi =
{1 \over \sqrt{L}}\Tr \Phi = {1 \over \sqrt{L}}\sum_{i=1}^{L}{\tilde \pi_i}$).
So $\lambda_{L-1}$ is the squared mass $m^2_{\phi_1}$ of the field $\phi_1$ 
and $\lambda_L$ is the squared mass $m^2_{\phi_2}$ of the field $\phi_2$.
At the leading order in the $N_c \to \infty$ limit, they are given by:
\ba
m^2_{\phi_1} &=& {2LA \over F_\pi^2 + LF_X^2} , \nonumber \\
m^2_{\phi_2} &=& {2c(F_\pi^2 + LF_X^2) \over F_\pi^2 F_X^2} .
\label{EQN5.13}
\ea
Let us observe that $m^2_{\phi_1}$ is of order ${\cal O}({1 \over N_c})$,
while $m^2_{\phi_2}$ is of order ${\cal O}(1)$.
As a check, we immediately see that, if we put $F_X=0$ in the above 
formulae (i.e., if we neglect the new chiral condensate), then $\phi_1 = S_\pi$
and $m^2_{\phi_1}$ reduces to ${2LA \over F^2_\pi}$, which is the ``usual''
$\eta'$ mass in the chiral limit \cite{Witten79a,Veneziano79}.
Moreover $m^2_{\phi_2}$ goes to infinity when
$F_X$ goes to zero, so that the field $\phi_2 = S_X$ is forced to be zero.
Yet, in the general case $F_X \ne 0$, the two states which diagonalize the 
squared mass matrix {\bf A} are linear combinations of the
``quark--anti-quark'' singlet field $S_\pi$ and of the ``exotic''
$2L$--fermion field $S_X$.
One of them ($\phi_1$) has a ``light'' mass, in the sense of the
$N_c \to \infty$ limit, being $m^2_{\phi_1} = {\cal O}({1 \over N_c})$;
this mass is intimately related to the anomaly and they both vanish in the 
$N_c \to \infty$ limit (see the first Eq. (\ref{EQN5.13})).
On the contrary the field $\phi_2$ has a sort of heavy ``hadronic'' mass
of order ${\cal O}(1)$ in the large--$N_c$ limit (see Ref. \cite{Witten79b}
for a detailed discussion about hadrons in the $1/N_c$ expansion).
Both the $\phi_1$ and the $\phi_2$ have the same quantum numbers (spin, 
parity and so on), but they have a different quark content: one is mostly
$S_\pi \sim i\sum_{i=1}^L{(\bar{q}_{iL}q_{iR}-\bar{q}_{iR}q_{iL})}$,
the other is mostly
$S_X \sim i[ {\det}(\bar{q}_{sL}q_{tR}) - {\det}(\bar{q}_{sR}q_{tL})]$.

What happens when approaching the chiral transition temperature $T_{ch}$?
We know that $F_\pi(T) \to 0$ when $T \to T_{ch}$. From the first
Eq.  (\ref{EQN5.13}) we see that $m^2_{\phi_1}(T_{ch}) = {2A \over F_X^2}$,
which is just the squared mass $m^2_{S_X}$ of the field $S_X$ for $T>T_{ch}$:
and in fact, from the first Eq. (\ref{EQN5.14}), $\phi_1(T_{ch}) = S_X$.
We have continuity in the mass spectrum of the theory through the chiral 
phase transition at $T=T_{ch}$. If we assume that $c(T)$ goes to zero at 
least as $F^2_\pi(T)$ when $T \to T_{ch}$, then also $m^2_{\phi_2}$ has a 
finite limit when $T \to T_{ch}$: $m^2_{\phi_2} \to
{2Lc \over F^2_\pi}$.

Another relevant comment has to be made about the field $\phi_1$ and its 
squared--mass formula, given by the first Eq. (\ref{EQN5.13}). It turns out
that $\phi_1$ is just the meson state, with a mass squared of order $1/N_c$, 
whose contribution to the full topological susceptibility $\chi$ exactly 
cancels out (in the chiral limit of massless quarks!) the pure gauge
part $A$ of $\chi$, so making $\chi = 0$: this is the so--called
Witten's mechanism, which is explained in Section 4.
Let us see how this picture comes out in our theory.
Let us determine, first, the $U(1)$ axial current, starting from our 
Lagrangian (\ref{EQN3.15}). This is easily done, remembering how the fields
$U$ and $X$ transform under a $U(1)$ chiral transformation
(see Eqs. (\ref{EQN3.7}) and (\ref{EQN3.10})).  We thus find the
following expression for the $U(1)$ axial current $J^{(L)}_{5, \mu}$:
\be
J^{(L)}_{5, \mu} = i[\Tr(U^\dagger \partial_\mu U - U \partial_\mu U^\dagger)
+ L(X^\dagger \partial_\mu X - X \partial_\mu X^\dagger)] .
\label{EQN5.15}
\ee
After having inserted here the expressions (\ref{EQN5.3}) and (\ref{EQN5.5})
in place of $U$ and $X$ respectively, the current $J^{(L)}_{5, \mu}$ takes the
following form:
\be
J^{(L)}_{5, \mu} = -\sqrt{2L} \partial_\mu (F_\pi S_\pi + \sqrt{L} F_X S_X) .
\label{EQN5.16}
\ee
In the case in which $S_X = 0$ (i.e., if we only have the usual chiral 
condensate $\langle \bar{q} q \rangle$), Eq. (\ref{EQN5.16}) reduces to
$J^{(L)}_{5, \mu} = -\sqrt{2L} F_S \partial_\mu S_\pi$,
where $F_S = F_\pi$ is the well--known expression
\cite{Witten79a,Veneziano79} for the singlet ($S_\pi$) decay constant, 
at the leading order in the $1/N_c$ expansion. Instead, in the general case
we are considering, the first Eq. (\ref{EQN5.14}) allows us to write
Eq. (\ref{EQN5.16}) as:
\be
J^{(L)}_{5, \mu} = -\sqrt{2L} F_{\phi_1} \partial_\mu \phi_1 .
\label{EQN5.17}
\ee
(where $F_{\phi_1}$ is defined below, in Eq. (\ref{EQN5.18})).
This means that the axial current $J^{(L)}_{5, \mu}$ only couples to the
``{\it light}'' field $\phi_1$, and not to the ``{\it heavy}'' field $\phi_2$
(``{\it light}'' and ``{\it heavy}'' are in the sense of $N_c$--order).
The relative coupling between $J^{(L)}_{5, \mu}$ and $\phi_1$, i.e., the
singlet ($\phi_1$) decay constant defined as
$\langle 0|J^{(L)}_{5, \mu}(0)|\phi_1(p) \rangle = i \sqrt{2L}p_\mu
F_{\phi_1}$, is now given by:
\be
F_{\phi_1} = \sqrt{F^2_\pi + LF^2_X} .
\label{EQN5.18}
\ee
Let us now recall the Witten's argument, which is discussed in Section 4,
and write the two--point function (at four--momentum $q$) of the 
topological charge density $Q(x)$ as a sum over one--hadron poles, i.e.,
one--hadron intermediate states:
\be
\chi(q) = -i\int d^4x e^{iqx} \langle TQ(x)Q(0) \rangle = A_0(q)
+ \displaystyle\sum_{mesons}{|\langle 0|Q|n \rangle |^2 \over q^2 - m^2_n} ,
\label{EQN5.19}
\ee
where $A_0(q)$ is the pure Yang--Mills contributions from the glueball
intermediate states and it is the leading--order term in $1/N_c$.
In the chiral limit in which we have $L$ massless quarks, the full topological
susceptibility $\chi \equiv \chi(q=0)$ must vanish, for the reasons explained
in Sections 4 and 5: so there has to be a meson state, with squared mass
$m^2_n = {\cal O}(1/N_c)$ (since $A_0(0)={\cal O}(1)$,
while $|\langle 0|Q|n \rangle |^2 = {\cal O}(1/N_c)$), which exactly cancels
out $A_0(0)$. This is the meson we usually call $\eta'$: the other meson states
in (\ref{EQN5.19}) have squared masses of order ${\cal O}(1)$, so that their
contributions to the summation in (\ref{EQN5.19}) are suppressed by a factor
of $1/N_c$. Therefore we obtain that:
\be
{|\langle 0|Q|\eta' \rangle |^2 \over m^2_{\eta'}} = A ,
\label{EQN5.20}
\ee
where $A \equiv A_0(0)$ is the pure Yang--Mills topological susceptibility
in the large--$N_c$ limit.
It is well-known that, in the chiral limit of massless quarks, the 
topological charge density $Q(x)$ is directly connected with the 
four--divergence of the axial current $J^{(L)}_{5, \mu}$, {\it via} the anomaly 
equation:
\be
\partial^\mu J^{(L)}_{5, \mu}(x) = 2L Q(x) ,
\label{EQN5.21}
\ee
so that $\langle 0|Q|\eta' \rangle = {1 \over \sqrt{2L}}m^2_{\eta'} F_{\eta'}$,
which can be substituted into Eq. (\ref{EQN5.20}) to give:
\be
A = {m^2_{\eta'} F^2_{\eta'} \over 2L} .
\label{EQN5.22}
\ee
Eq. (\ref{EQN5.22}) relates the mass $m_{\eta'}$ of the $\eta'$ state, its
decay constant $F_{\eta'}$ and the pure--gauge topological susceptibility $A$.
It is easy to see, now, that the $\phi_1$ state, whose mass and decay 
constant are given respectively by Eqs. (\ref{EQN5.13}) and (\ref{EQN5.18}),
really satisfies Eq. (\ref{EQN5.22}): it is just the $\eta'$ state!

One could have also followed an inverse reasoning, starting from Witten's 
assumption that there has to be a meson state, called $\eta'$, with squared
mass $m^2_{\eta'} = {\cal O}(1/N_c)$, which couples to $J^{(L)}_{5, \mu}$ in
such a way to exactly cancel $A = A_0(0)$ in Eq. (\ref{EQN5.19}).
Looking at Eq. (\ref{EQN5.16}) and comparing it with the first
Eq. (\ref{EQN5.14}), it seems natural to conclude that (doing also a 
suitable state--normalization) the state $\eta'$ is nothing but $\phi_1$, 
with a decay constant given by Eq. (\ref{EQN5.18}). One thus derives the
$\phi_1$ mass from Eq. (\ref{EQN5.22}), which obviously furnishes the usual
value (\ref{EQN5.13}):
$m^2_{\phi_1} = {2LA \over F^2_{\phi_1}} = {2LA \over F_\pi^2 + LF_X^2}$.
Therefore our theory seems to be self--consistent.

\newsection{The real-world case at $T<T_{ch}$}

\noindent
Let us apply, now, the results of the previous Section to the
``{\it real--world}'' case, in which there are $L = 3$ light flavours, 
named $u$, $d$ and $s$, with masses $m_u$, $m_d$, $m_s$ small compared to 
the QCD mass--scale $\Lambda_{QCD}$ (or better: $m_u,~m_d \ll m_s \ll
\Lambda_{QCD}$) \cite{EM1994c}.
It is useful to represent $\Phi$ as $\Phi = \sum_{a = 1}^8
\pi_a \tau_a + {S_\pi \over \sqrt{3}}{\bf I}$,
$\tau_a$ ($a=1,\ldots,8$) being the generators of the algebra of $SU(3)$
in the fundamental representation, with normalization:
\be
\Tr(\tau_a) = 0 ~~~ ; ~~~ \Tr(\tau_a \tau_b) = \delta_{ab}.
\label{EQN6.1}
\ee
(In other words $\tau_a = \lambda_a /\sqrt{2}$, where $\lambda_a$ are the 
Gell--Mann matrices.)
The fields $\pi_a$ describe the mesons of the octet, while $S_\pi$ is an
$SU(3)$--singlet field.  In terms of $\Phi$ and $S_X$, the quadratic part of
the Lagrangian (\ref{EQN3.15}) reads:
\ba
\lefteqn{
{\cal L}_2 = {1 \over 2}\Tr\left( {\partial_\mu \Phi \partial^\mu \Phi} \right)
+{1 \over 2}\partial_\mu S_X \partial^\mu S_X 
-{B_m \over 2F_\pi}\Tr \left[ M \Phi^2 \right] + }
\nonumber \\
& & -c \left( {1 \over F_\pi} \Tr\Phi - {1 \over F_X}S_X \right) ^2 
-A\left( {\omega_1 \over F_\pi} \Tr\Phi
+ {1-\omega_1 \over F_X}S_X \right) ^2 .
\label{EQN6.2}
\ea
If we introduce in place of $\Phi$ its expression in terms of the fields
$\pi_a$ and $S_\pi$,
we immediately find that the states $\pi_1,~\pi_2,~\pi_4,~\pi_5,~\pi_6,~
\pi_7$ are already diagonal, with masses:
\ba
m_{\pi_{1,2}}^2 &\equiv& m_{\pi^\pm}^2 = B(m_u + m_d), \nonumber \\
m_{\pi_{4,5}}^2 &\equiv& m_{K^\pm}^2 = B(m_u + m_s), \nonumber \\
m_{\pi_{6,7}}^2 &\equiv& m_{K^0,\bar{K}^0}^2 = B(m_d + m_s),
\label{EQN6.3}
\ea
where we have put $B \equiv {B_m \over 2F_\pi}$.
On the contrary the states $\pi_3,~\pi_8,~S_\pi,~S_X$ mix together.
However, if we neglect terms of order ${\cal O}({(m_u-m_d) \over
\Lambda_{QCD}})$, as it is usually done in chiral perturbation theory 
(they are effects of $SU(2)$ 
isospin breaking, which are experimentally very small),
we find that also the state $\pi_3$ becomes ``diagonal'', with mass:
\be
m_{\pi_{3}}^2 \equiv m_{\pi^0}^2 = B(m_u + m_d) = 2B\tilde{m},
\label{EQN6.4}
\ee
where we have put: $\tilde{m} \equiv {m_u + m_d \over 2}$.

The interesting results are obviously obtained when considering the system 
of fields $\pi_8,~S_\pi,~S_X$, having the following squared mass matrix:
\be
{\bf A} =\pmatrix{
{2B{\tilde{m}+2m_s \over 3}} & {2B{\sqrt{2} \over 3}(\tilde{m}-m_s)} &
{0} \cr
{2B{\sqrt{2} \over 3}(\tilde{m}-m_s)} & 
{6(A\omega_1^2 +c) \over F_\pi^2} + f_0 & 
{2A\omega_1(1-\omega_1)-2c \over F_\pi F_X} \cr
0 & {2A\omega_1(1-\omega_1)-2c \over F_\pi F_X} & 
{2A(1-\omega_1)^2 +2c \over F_X^2} \cr} ,
\label{EQN6.5}
\ee
where $f_0 \equiv {2 \over 3}B(m_u + m_d + m_s) = {2 \over 3}
B(2\tilde{m} + m_s)$.
The eigenvalues of {\bf A} may be quite easily obtained at the first order 
in $m_i$ and $1/N_c$ (so that, for example, terms of order ${\cal O}(m_i/N_c)$
are considered of second order). 
For this purpose we need to know the $N_c$--dependences of the various 
quantities appearing in the matrix {\bf A}. They will be derived in
Section 14, with the result:
\be
F_\pi = {\cal O}(N_c^{1/2});~~ F_X = {\cal O}(N_c^{1/2});~~ A = {\cal O}(1);~~
c = {\cal O}(N_c).
\label{EQN6.6}
\ee
The eigenvalues of {\bf A} are found to be:
\ba
m_\eta^2 &=& 2B {\tilde{m} + 2m_s \over 3}, \nonumber \\
m_{\eta'}^2 &=& {6A \over F_\pi^2 + 3F_X^2} +{F_\pi^2 \over F_\pi^2 + 
3F_X^2} \cdot {2 \over 3}B(2\tilde{m} + m_s), \nonumber \\
m_{\eta_X}^2 &=& {2c(F_\pi^2 + 3F_X^2) \over F_\pi^2 F_X^2} +
{\cal O}_1({1 \over N_c}) + {3F_X^2 \over F_\pi^2 + 3F_X^2} \cdot {2 \over 3}B
(2\tilde{m} + m_s) ,
\label{EQN6.7}
\ea
where ${\cal O}_1(1/N_c)$ is a quantity of order $1/N_c$, which also depends
on the parameter $\omega_1$ appearing in the Lagrangian (\ref{EQN6.2}).
The physical interpretation of these three states is clear. The $\eta$--state
is the eighth pseudo--Goldstone meson of the octet: its mass vanishes with the
(light) quark masses. On the contrary the $\eta'$--state and the
$\eta_X$--state have masses which do not vanish when the (light) quark masses
are put equal to zero.
Yet, while the $\eta'$--state has a ``non--chiral'' topological mass
${6A \over F_\pi^2 + 3F_X^2}$, which is of order ${\cal O}(1/N_c)$ in the
$1/N_c$ expansion, the $\eta_X$--state has a sort of heavy ``hadronic'' mass
of order ${\cal O}(1)$ in the large--$N_c$ limit.
Both the $\eta'$ and the $\eta_X$ have the same quantum numbers (spin, 
parity and so on), but they have a different quark content: one is mostly
$\sim i\sum_{i=1}^3{(\bar{q}_{iL}q_{iR}-\bar{q}_{iR}q_{iL})}$, the other is
mostly $\sim i[ {\det}(\bar{q}_{sL}q_{tR}) - {\det}(\bar{q}_{sR}q_{tL})]$.
From Eqs. (\ref{EQN6.3}), (\ref{EQN6.4}) and (\ref{EQN6.7}) we find that
the squared masses of the octet mesons satisfy the Gell-Mann--Okubo (GMO)
formula, which is a standard result of chiral perturbation theory at the
lowest order:
\be
3m_\eta^2 + m_\pi^2 = 4m_K^2,
\label{EQN6.8}
\ee
considering: $m_u \simeq m_d \simeq \tilde{m}$. In fact it is natural 
to suppose that the introduction of a new chiral order parameter, which 
breaks only the $U(1)$ axial symmetry, should not modify the mass relations
(such as the (\ref{EQN6.8})) of the $SU(3)$ meson octet: these relations only
derive from the breaking of $SU(3) \otimes SU(3)$ down to $SU(3)_V$.

The new interesting result is obtained considering also the second equation 
of (\ref{EQN6.7}). From this we immediately derive the following expression
\cite{EM1994c}:
\be
\left( 1+3{F_X^2 \over F_\pi^2} \right) m_{\eta'}^2 + m_\eta^2
-2m_K^2 = {6A \over F_\pi^2}.
\label{EQN6.9}
\ee
Eq. (\ref{EQN6.9}) is an obvious and natural generalization of the
Witten--Veneziano formula for the $\eta'$, derived in Refs.
\cite{Witten79a,Veneziano79}.
Now the presence of the new $U(1)$ chiral order parameter induces a 
correction of order ${F_X^2 / F_\pi^2}$, where $F_X$ is essentially the 
magnitude of this chiral condensate. It is a beautiful and interesting 
thing that Eq. (\ref{EQN6.9}) does not contain the other (unknown) parameters
of our model (such as $c_1,~\omega_1, \ldots$).
One may ask if this is true only for our simplified model. In the next 
Section we will see that the result (\ref{EQN6.9}) is really
``model--independent'', being still valid when one adds to our ``simple'' 
model other possible additional terms.

\newsection{The effects of additional terms in the Lagrangian}

\noindent
As we have said in Section 9, the form (\ref{EQN3.13})--(\ref{EQN3.14}) for the
chiral effective Lagrangian, at the leading order in $m_i$ and $1/N_c$, is
surely the most simple and natural one, but not the most general one.
Now we want to see if our result (\ref{EQN6.9}) is
``model--independent'', and we will find that this is just the case.

To prove this, we first consider the effects produced by an additional
term, linear in the quark mass matrix $M$, of the form:
\ba
\lefteqn{
\delta {\cal L}^{(2)}_{mass} =
\rho_1 \left[ \Tr(MU)\det(U)X^\dagger + \Tr(M^\dagger
U^\dagger )\det(U^\dagger )X \right] + }
\nonumber \\
& & + \rho_2 \left[ \Tr(MU)\det(U^\dagger )X + \Tr(M^\dagger U^\dagger )
\det(U)X^\dagger \right] .
\label{EQN7.1}
\ea
This new term may also be written in the following form:
\ba
\lefteqn{
\delta {\cal L}^{(2)}_{mass} =
S \left[ \Tr(MU)+\Tr(M^\dagger U^\dagger) \right] \cdot
\left[ \det(U)X^\dagger + \det(U^\dagger )X \right] + }
\nonumber \\
& & + D \left[ \Tr(MU)-\Tr(M^\dagger U^\dagger) \right] \cdot
\left[ \det(U)X^\dagger - \det(U^\dagger )X \right] ,
\label{EQN7.2}
\ea
where the new parameters $S$ and $D$ are linked to $\rho_1$ and $\rho_2$
by the following relation:
\be
S = {\rho_1 + \rho_2 \over 2} ~~~,~~~ D = {\rho_1 - \rho_2 \over 2}.
\label{EQN7.3}
\ee
Proceeding as in the previous Section, we find that the squared masses of 
the pseudoscalar mesons are given by (expanding to the first order in
$m_i$ and $1/N_c$):
\ba
m_{\pi_{1,2,3}}^2 &\equiv& m_\pi^2 = B'(m_u + m_d),
\nonumber \\
m_{\pi_{4,5}}^2 &\equiv& m_{K^\pm}^2 = B'(m_u + m_s),
\nonumber \\
m_{\pi_{6,7}}^2 &\equiv& m_{K^0,\bar{K}^0}^2 = B'(m_d + m_s),
\nonumber \\
m_\eta^2 &=& 2B' {\tilde{m} + 2m_s \over 3},
\nonumber \\
m_{\eta'}^2 &=& {6A \over F_\pi^2 + 3F_X^2} +{F_\pi^2 \over F_\pi^2 + 
3F_X^2} \cdot {2 \over 3}B'(2\tilde{m} + m_s),
\nonumber \\
m_{\eta_X}^2 &=& {2c'(F_\pi^2 + 3F_X^2) \over F_\pi^2 F_X^2} + {\cal O}_1(
{1 \over N_c}) + {3F_X^2 \over F_\pi^2 + 3F_X^2} \cdot {2 \over 3}B'
(2\tilde{m} + m_s)
\nonumber \\
~~~ &+& 2\sqrt{2} F_\pi^2 F_X D (2\tilde{m} + m_s),
\label{EQN7.4}
\ea
where the new parameters $B'$ and $c'$ are given by:
\ba
B' &=& B + {\sqrt{2} \over 2} F_\pi^2 F_X S, \nonumber \\
c' &=& c + {\sqrt{2} \over 2} F_\pi^4 F_X S (2\tilde{m} + m_s).
\label{EQN7.5}
\ea
Observing Eqs. (\ref{EQN7.4}), we immediately see that both the GMO
relation (\ref{EQN6.8}) and our {\it generalized Witten--Veneziano formula}
(\ref{EQN6.9}) are still valid. The effect of the introduction of the
additional term (\ref{EQN7.2}) to the effective 
Lagrangian (\ref{EQN3.13})--(\ref{EQN3.14}) on the masses of the octet mesons
and of the $\eta'$ is therefore only that of rescaling the parameter 
$B \to B' = B + {\sqrt{2} \over 2} F_\pi^2 F_X S$ (note that only the 
parameter $S$ appears here): the relations (\ref{EQN6.8}) and (\ref{EQN6.9})
do not depend (in form) on the value of B, so that they remain valid.
In a certain sense, the ``model--dependence'' is absorbed entirely by the 
squared mass $m_{\eta_X}^2$ of the $\eta_X$--state, which is ``large'' in 
the sense of the $1/N_c$ expansion, being of order ${\cal O}(1)$.

By a similar analysis, it is not hard to verify that also the 
introduction of other (more complicated) additional terms, linear in
$M$, has the same effect: it simply rescales $S$ and $D$ in
Eqs. (\ref{EQN7.4}), and the relations (\ref{EQN6.8}) and (\ref{EQN6.9})
are still valid.

One may also try to generalize the last term in the potential (\ref{EQN3.14})
by adding, for example, a term of the form:
\be
\delta V = -{c_2 \over 8\sqrt{2}} \left[ \det(U)X^\dagger 
+ \det(U^\dagger )X \right]^2 .
\label{EQN7.6}
\ee
It is immediate to see that its only effect on the quadratic part ${\cal L}_2$
of the Lagrangian written in Eq. (\ref{EQN6.2}) is to re--scale the parameter
$c$:
\be
c \rightarrow c + {c_2 \over \sqrt{2}} \left( {F_X \over \sqrt{2}} \right)^2
\left( {F_\pi \over \sqrt{2}} \right)^6 .
\label{EQN7.7}
\ee
One can easily convince oneself that a general modification of the 
above--mentioned term in the potential $V(U,U^\dagger ,X,X^\dagger)$
(Eq. (\ref{EQN7.6}) is an example of such a modification) implies only a
re--scaling of the parameter $c$ in the quadratic Lagrangian (\ref{EQN6.2}).
Once again all the ``model--dependence'' is absorbed by the 
squared mass $m_{\eta_X}^2$ of the $\eta_X$--state.

The GMO formula (\ref{EQN6.8}) and the {\it generalized Witten--Veneziano
formula} (\ref{EQN6.9}) acquire the aspect of theorems: they are not sensitive
to the specificity of the model. 
And while the GMO formula does not depend on the new $U(1)$ chiral order 
parameter (as it must be!), we find that the Witten--Veneziano formula
(which links together the $\eta'$ mass, the $\eta$ and $K$ masses and the
pure--gauge topological susceptibility A) is modified through a term which 
only contains $F_X$, i.e., essentially the magnitude of this $U(1)$ 
chiral order parameter. Experimentally we have that:
\ba
m_{K^\pm} &=& ~493.646 \pm 0.009 ~{\rm MeV},
\nonumber \\
m_{K^0, \bar{K}^0} &=& ~497.671 \pm 0.0031 ~{\rm MeV},
\nonumber \\
m_{\eta} &=& ~547.45 \pm 0.19 ~{\rm MeV},
\nonumber \\
m_{\eta'(958)} &=& ~957.75 \pm 0.14 ~{\rm MeV}.
\label{EQN7.8}
\ea
From lattice simulations for the $SU(3)$ Yang--Mills theory (QCD with no 
quarks) one obtains the following value for the topological susceptibility A
(see, e.g., Refs. \cite{Teper88,Campostrini-SU(3)} and \cite{Alles-et-al.97}):
\be
A \simeq (179 \pm 4 ~{\rm MeV})^4.
\label{EQN7.9}
\ee
Then Eq. (\ref{EQN6.9}) implies that, within the present errors, a value for
$F_X$ different from zero is not excluded, up to a superior limit of about 
$10 \div 20$ MeV (we remind, for a comparison, that the pion decay constant
is about $F_\pi \simeq 93$ MeV).
Of course, one should also remark that Eq. (\ref{EQN6.9}) has been derived
in the large--$N_c$ limit, and possible next--to--leading corrections
in the $1/N_c$ expansion could be non--negligible at the physical value
$N_c = 3$ (see, for example, Ref. \cite{Peris94}), so becoming comparable
with (or even larger than) the (presumably!) small correction
$\sim F_X^2 / F_\pi^2$, induced by the new $U(1)$ chiral condensate.

\newsection{The topological susceptibility and the chiral condensate}

\noindent
After having discussed in the previous Sections the mass spectrum of the 
theory above and below the chiral phase transition at $T=T_{ch}$, we can 
now see which are the predictions of our model about the topological 
susceptibility and the chiral condensate. Let us begin with the topological 
susceptibility, defined as $\langle TQ(x)Q(0) \rangle^{F.T.}(k=0)$,
where ``$F.T.$'' stands for the Fourier transformed.
If we want to derive this two--point 
function of $Q(x)$, we need to consider the Lagrangian in the form 
(\ref{EQN3.13}), where the field variable $Q(x)$ has not yet been integrated.
So doing, we obtain the following expression for the two--point function of
$Q(x)$, in the region of temperatures $T_{ch} < T < T_{U(1)}$:
\be
\langle TQ(x)Q(0) \rangle^{F.T.}(k) = i({\bf A}^{-1}(k))_{11} ,
\label{EQN8.1}
\ee
where ${\bf A}^{-1}(k)$ is the inverse of the following squared mass matrix
for the ensemble of fields $(Q(x),~S_X,~b_{11},~b_{12}, \ldots )$
(see Section 10):
\be
{\bf A}(k) =\pmatrix{
{1 \over A} & -{\sqrt{2} \over F_X} & 0 & \ldots \cr
-{\sqrt{2} \over F_X} & k^2-m^2_0 & {\cal O}(m^{L-1}) & \ldots \cr
0 & {\cal O}(m^{L-1}) & k^2 -{1 \over 2} \lambda_\pi^2 B_\pi^2 & \ldots \cr
\vdots & \vdots & \vdots & \ddots \cr} ,
\label{EQN8.2}
\ee
where $m^2_0$ is the eigenvalue proportional to $\det(M)= \prod_{k=1}^L m_k$,
written explicitly in Eq. (\ref{EQN4.3}). Doing explicitly 
the calculations, we obtain that:
\be
\langle TQ(x)Q(0) \rangle^{F.T.}(k) = i({\bf A}^{-1}(k))_{11} =
iA{k^2-m^2_0 \over k^2 - m^2_{S_X} } ,
\label{EQN8.3}
\ee
where, as usual: $m^2_{S_X} = m^2_0 + {2A \over F_X^2}$.
Therefore the topological susceptibility in this region $T_{ch} < T < 
T_{U(1)}$, is given by:
\ba
\lefteqn{
\langle TQ(x)Q(0) \rangle^{F.T.}(k=0) = iA{m^2_0 \over m^2_{S_X} } =
iA{m^2_0 \over m^2_0 +{2A \over F_X^2}} }
\nonumber \\
& &\mathop{\simeq}_{\sup(m_i) \to 0} i{F_X^2 \over 2} m^2_0 =
i{1 \over 2}c_1 F_X \left( {2B_m \over \sqrt{2} \lambda_\pi^2 B_\pi^2} 
\right)^L \det(M) .
\label{EQN8.4}
\ea
So we obtain that the topological susceptibility of the full theory (not of
the pure YM theory!) in the region $T_{ch} < T < T_{U(1)}$ is proportional to
the product of the light quark masses $m_i$ ($i=1,2,\ldots ,L$). This is the
same behaviour that can be derived from a simple instantonic model,
as shown in the Appendix.
We also find that the propagator of the field $S_X$ is given by:
\be
\langle TS_X(x)S_X(0) \rangle^{F.T.}(k) =
i({\bf A}^{-1}(k))_{22}={i \over k^2- m^2_{S_X}} ,
\label{EQN8.5}
\ee
which is consistent with what was found in Section 10: i.e., the 
field $S_X$ has a squared mass $m^2_{S_X} = m^2_0 + {2A \over F_X^2}$.

What happens in the region $T<T_{ch}$? Here we have to consider the 
following quadratic Lagrangian, including the field $Q(x)$ (see Sections
9 and 11):
\ba
\lefteqn{
{\cal L}_2 ={1 \over 2}\displaystyle\sum_{i=1}^L{\partial_\mu \tilde\pi_i
\partial^\mu \tilde\pi_i} 
+{1 \over 2}\displaystyle\sum_{\alpha \ne \beta}{\partial_\mu \tilde\pi^
{\alpha \beta} \partial^\mu \tilde\pi^{\alpha \beta}} 
+{1 \over 2}\partial_\mu S_X \partial^\mu S_X + }
\nonumber \\
& & -{1 \over 2}\displaystyle\sum_{i=1}^L {\mu_i^2 \tilde\pi_i^2}
-{1 \over 2}\displaystyle\sum_{\alpha \ne \beta}{{\mu_\alpha^2 +\mu_\beta^2 
\over 2} \tilde\pi^{\alpha \beta} \tilde\pi^{\alpha \beta}} +
\nonumber \\
& & -{1 \over 2}c\left( {\sqrt{2} \over F_\pi}\displaystyle\sum_{i=1}^L
{\tilde\pi_i}
-{\sqrt{2} \over F_X}S_X \right) ^2 +{1 \over 2A} Q^2(x) +
\nonumber \\
& & -\omega_1{\sqrt{2} \over F_\pi}Q(x)\displaystyle\sum_{i=1}^L{\tilde\pi_i}
- (1-\omega_1) { \sqrt{2} \over F_X}Q(x)S_X .
\label{EQN8.6}
\ea
Proceeding as in the previous case, we derive the following expression (in 
the special case $L=2$) for the topological susceptibility in the region
$T<T_{ch}$, below the chiral phase transition:
\be
\langle TQ(x)Q(0) \rangle^{F.T.}(k=0)=iA{ \mu^2_1 \mu^2_2 \over
{A(1-\omega_1)^2+c \over c}\mu^2_1 \mu^2_2 +{2A \over F_\pi^2}
(\mu^2_1 + \mu^2_2)} ,
\label{EQN8.7}
\ee
which has the usual dependence on the light quark masses in this 
region: it goes as $i{F_{\pi}^2 \over 2}\mu_l^2 \propto m_l$ for $m_l \to 
0$ (Using Eq. (\ref{EQN8.9}), that we will derive below, and remembering that
$\mu^2_l = {B_m \over F_\pi}m_l$, we derive that 
$\langle TQ(x)Q(0) \rangle^{F.T.}(k=0)=-i m_l \langle \bar{q}_l q_l \rangle$,
for $T<T_{ch}$ and $m_l \to 0$, as requested by the QCD Ward Identities at the
leading order in the light quark masses \cite{Crewther79}).
 
Now we address the question of the chiral condensate. It is well known that 
the derivative of the QCD Hamiltonian with respect to $m_i$ is the operator
$\bar{q}_i q_i$ (being $\delta {\cal L}_{QCD}^{(mass)} =
-\sum_{i=1}^L{m_i \bar{q}_i q_i}$).
The corresponding derivative of the vacuum energy 
therefore represents the vacuum expectation value of $\bar{q}_i q_i$.
So, in our case, it must be:
\be
\langle \bar{q}_i q_i \rangle = {\partial \over \partial m_i}
\langle V(U,U^\dagger ,X,X^\dagger ) \rangle ,
\label{EQN8.8}
\ee
where $V(U,U^{\dagger},X,X^{\dagger})$ is the potential term written in
Eq. (\ref{EQN3.14}). It is immediate to calculate the expectation value of the 
potential V in the two regions $T<T_{ch}$ and $T_{ch} < T < T_{U(1)}$.
In the region $T<T_{ch}$ we find that (always at the leading order in
$m_i$, $1/\lambda_{\pi}^2$, $1/\lambda_X^2$):
\be
\langle \bar{q}_i q_i \rangle_{T<T_{ch}} \simeq -{1 \over 2}B_m F_\pi .
\label{EQN8.9}
\ee
It is of order one in the light quark masses, indicating 
that the $SU(L) \otimes SU(L)$ chiral symmetry is really broken.
Of course the interesting result is obtained in the region $T_{ch} < T < 
T_{U(1)}$, above the $SU(L) \otimes SU(L)$ chiral transition. Here we find that:
\be
\langle \bar{q}_i q_i \rangle_{T_{ch}<T<T_{U(1)}} \simeq 
-{B_m^2 \over \lambda_\pi^2 B_\pi^2}m_i -{1 \over 2}c_1 F_X 
\left( {2B_m \over \sqrt{2} \lambda_\pi^2 B_\pi^2} \right)^L 
(\displaystyle\prod_{k \ne i}{m_k} ) .
\label{EQN8.10}
\ee
The interpretation of the two terms in the right--hand side of
Eq. (\ref{EQN8.10}) is rather simple in a diagrammatic language.
The first term, linear in the mass $m_i$, corresponds to a diagram
with a mass insertion, as shown in Fig. 5.

\begin{figure}[htb]
\vskip 4.5truecm
\includegraphics{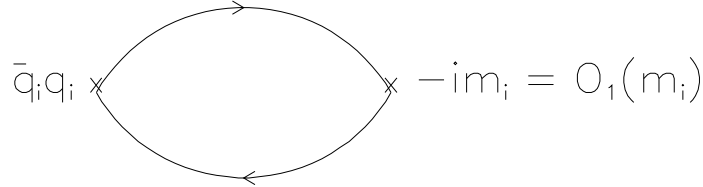}
\caption{The first term in the RHS of Eq. (\ref{EQN8.10}).}
\end{figure}

\noindent
Instead the second term clearly corresponds to an insertion of the new 
$2L$--fermion effective vertex associated with $X$, as shown in Fig. 6.

\begin{figure}[htb]
\vskip 4.5truecm
\includegraphics{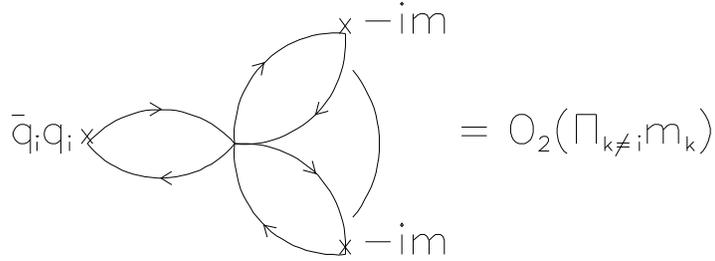}
\caption{The second term in the RHS of Eq. (\ref{EQN8.10}).}
\end{figure}

\noindent
In conclusion we have obtained that, in the region $T_{ch} < T < T_{U(1)}$, 
the chiral condensate $\langle \bar{q}_i q_i \rangle $ has the following
behaviour with respect to the light quark masses:
\be
\langle \bar{q}_i q_i \rangle =
{\cal O}_1(m_i) + {\cal O}_2(\displaystyle\prod_{k \ne i}{m_k}) .
\label{EQN8.11}
\ee
This relation can be derived also in a simple instantonic model, as shown 
in the Appendix. Comparing the explicit expression for
${\cal O}_2(\prod_{k \ne i} {m_k})$, 
written in the right--hand side of Eq. (\ref{EQN8.10}), 
with the expression for the full topological susceptibility given by 
Eq. (\ref{EQN8.4}), we find the following interesting relation:
\be
\langle TQ(x)Q(0) \rangle^{F.T.}(k=0) =
-i m_l{\cal O}_2(\displaystyle\prod_{k \ne l}{m_k}),
\label{EQN8.12}
\ee
which is, of course, independent of the flavour index ``$l$''.

We want to close this Section with some comments about the $N_c$--dependence
of some relevant quantities that we have considered in our analysis.
In Section 8 we have shown that the picture we have discussed so 
far is really consistent with the saturation of the QCD Ward Identities 
above the $SU(L) \otimes SU(L)$ chiral phase transition \cite{EM1994b}.
The saturation is obtained in the framework of the $1/N_c$ expansion, where 
$N_c$ is the number of colours and we derive the $N_c$ 
dependence of a certain number of relevant quantities.
In particular the squared mass $m^2_{S_X}$ of the pseudoscalar singlet boson in
the region $T_{ch}<T<T_{U(1)}$, and in the chiral limit of zero quark 
masses, depends on $N_c$ as ${\cal O}(1/N_c)$. From Eq. (\ref{EQN4.5}) we have
that $m^2_{S_X} = {2A \over F^2_X}$ for zero quark masses; considering that
$A={\cal O}(1)$ with respect to $N_c$ (see Refs. \cite{Witten79a,Veneziano79}),
we immediately derive that $F_X = {\cal O}(N_c^{1/2})$.
In Eq. (\ref{EQN8.9}) we have found that the chiral condensate in the region 
$T<T_{ch}$ is given by: 
$\langle \bar{q}_i q_i \rangle_{T<T_{ch}} \simeq -{1 \over 2}B_m F_\pi$.
It is well known (see Refs. \cite{Witten79a} and \cite{tHooft74,Witten79b})
that both $F^2_\pi$ and $\langle \bar{q}_i q_i \rangle$ are of order
${\cal O}(N_c)$ (and this remains true also considering the new condensate
which breaks the $U(1)_A$ symmetry).
This fact implies that $B_m = {\cal O}(N_c^{1/2})$.
In Section 8 we have also found that the two contributions ${\cal O}_1(m_i)$
and ${\cal O}_2(\prod_{k \ne i}{m_k})$ for the chiral condensate
$\langle \bar{q}_i q_i \rangle$ in the region $T_{ch}<T<T_{U(1)}$
(see Fig. 4 in Section 14) are both of order
${\cal O}(N_c)$. This means, considering the explicit expressions for
${\cal O}_1(m_i)$ and ${\cal O}_2(\prod_{k \ne i}{m_k})$
written in Eq. (\ref{EQN8.10}), that $\lambda^2_\pi B^2_\pi = {\cal O}(1)$
with respect to $N_c$ and that $c_1 = {\cal O}(N_c^{(1-L) / 2})$.
Therefore, from Eq. (\ref{EQN5.8}), $c = {\cal O}(N_c)$.
As a consequence of this, we have derived in Section 11 the 
$N_c$--dependence of the non-zero--mass states $\phi_1$ and $\phi_2$
(whose explicit expressions in terms of the fields $S_\pi$ and $S_X$ are 
given in Eq. (\ref{EQN5.14})) in the region $T<T_{ch}$. We have found that the
state $\phi_1$ has a ``light'' mass $m_{\phi_1}$, as $N_c \to \infty$, being
$m^2_{\phi_1} = {2LA \over F^2_\pi + LF^2_X} = {\cal O}({1 \over N_c})$.
Moreover $m_{\phi_1}$ is purely originated by the anomaly (it vanishes if
A is put equal to zero!). On the contrary the state $\phi_2$ has a sort of
heavy ``hadronic'' mass $m_{\phi_2}$ of order ${\cal O}(1)$ with respect to
$N_c$ and not dependent on the anomaly.

\newsection{Conclusions}

\noindent
In this paper (after a pedagogical review of the $U(1)$ problem at $T=0$,
which has been presented in Sections 2--5), we have tried to gain a physical
insight into the breaking mechanism of the $U(1)$ axial symmetry, through a
study of the behaviour of the theory at finite temperature.
As discussed in the Introduction, the topological susceptibility $A$ of the
pure Yang--Mills theory is a fundamental quantity for studying the $U(1)$
chiral symmetry, both at zero and non--zero temperature.
From some previous works of Witten \cite{Witten79a} and Veneziano
\cite{Veneziano79,Veneziano80}, it is known that a value for $A$ different 
from zero implies, at large number $N_c$ of colours, the breaking of the
$U(1)$ axial symmetry, since it implies the existence of a 
pseudo--Goldstone particle with the same quantum numbers of the $\eta'$.
In the Witten--Veneziano model the mass $m_{S_\pi}$ of the $SU(L)$--singlet
field for $T<T_{ch}$ acquires a ``topological'' contribution 
from the topological susceptibility in such a way that, in the 
chiral limit of zero quark masses, $m^2_{S_\pi}={2LA \over F_\pi ^2}$
($F_\pi$ is the usual pion decay constant: $F_\pi \simeq 93$ MeV at
$T=0$). Approaching the chiral phase transition temperature $T_{ch}$, 
$F_\pi$ vanishes and so we could expect that also $A$ goes to zero 
approaching $T_{ch}$, in order to avoid a singularity in the singlet field 
mass for $T \to T_{ch}$.
Yet, recent lattice results \cite{Alles-et-al.97} show that the YM topological
susceptibility $A(T)$ is approximately constant up to the critical temperature
$T_{ch}$, it has a sharp decrease above the transition, but it remains
different from zero up to $\sim 1.2~T_{ch}$.
Also present lattice data for the so--called ``chiral susceptibilities''
(discussed in the Introduction) seem to indicate that the $U(1)$ axial
symmetry is still broken above $T_{ch}$, up to $\sim 1.2~T_{ch}$.

In the following, we briefly summarize the main points
that we have discussed and the results that we have obtained.
\begin{itemize}
\item{
One expects that, above a certain critical temperature $T_{U(1)}$, also the
$U(1)$ axial symmetry will be (approximately) restored. We have tried to see
if this transition has (or has not) anything to do with the usual
$SU(L) \otimes SU(L)$ chiral transition: various possible scenarios
have been discussed in Section 6.
}
\item{
We have proposed a scenario (supported by the above--mentioned lattice
results) in which the $U(1)$ axial symmetry is still broken above the chiral
transition and the pure YM topological susceptibility vanishes at a
temperature $T_\chi$ between $T_{ch}$ and $T_{U(1)}$. A new order parameter
has been introduced in Section 7 for the $U(1)$ axial symmetry.
In Section 8  we have shown that this picture is consistent with the 
saturation of the QCD Ward Identities above the $SU(L) \otimes SU(L)$ 
chiral phase transition, in the framework of the $1/N_c$ expansion,
taking care also of the new $2L$--fermion condensate which breaks the
$U(1)_A$ symmetry.
}
\item{
In Sections 9--14 we have shown that this scenario can be 
consistently reproduced using an effective Lagrangian model.
We have analysed the effects that one should observe on the mass 
spectrum of the theory, both below and above $T_{ch}$. In particular, the
Witten--Veneziano formula for the mass of the $\eta'$ is modified by the
presence of the new $2L$--fermion condensate. In this scenario,
the $\eta'$ survives across the chiral transition at $T_{ch}$ in the form of
an ``exotic'' $2L$--fermion state
$\sim i[ {\det}(\bar{q}_{sL}q_{tR})$ $-{\det}(\bar{q}_{sR}q_{tL})]$.
This particle is nothing but the {\it would--be} Goldstone boson coming from
the breaking of the $U(1)$ axial symmetry.
For $T>T_{ch}$, it acquires a ``topological'' squared 
mass of the form $2A / F_X^2$, where $F_X$ is essentially the magnitude of
the new order parameter for the $U(1)$ chiral symmetry alone.
}
\end{itemize}
This scenario could perhaps be verified in the near future by heavy--ions 
experiments, by analysing the spectrum in the singlet sector.
If the ``exotic'' light state would be copiously produced at 
high $T$, its subsequent decay into $L$ light mesons (as the  
temperature $T$ of the plasma formed in the collision decreases), 
could be seen in terms of a production of $L$ quark--antiquark couples
$\bar{q}_i q_i$ ($i=1, \ldots ,L$) and their subsequent linear 
superpositions to form $L$ physical mesons. This is quite similar to high 
energy instanton effects.
Some tests and verifications of this picture could also be provided by Monte
Carlo simulations on the lattice. We hope that further progress along this
line will be done in the near future.

\vfill\eject

\renewcommand{\thesection}{}
\renewcommand{\thesubsection}{A.\arabic{subsection} }
 
\pagebreak[3]
\setcounter{section}{1}
\setcounter{equation}{0}
\setcounter{subsection}{0}
\setcounter{footnote}{0}

\begin{flushleft}
{\large\bf \thesection Appendix:
the quark mass dependence for $T>T_{ch}$}
\end{flushleft}

\renewcommand{\thesection}{A}


\noindent
In this Appendix we derive a (possible) quark mass dependence of the 
topological susceptibility $\chi$ of the full theory (QCD with quarks)
and of the chiral condensate $\langle \bar{q}_i q_i \rangle$ in the chirally
symmetric phase (i.e., for $T>T_{ch}$). We follow a procedure outlined in Ref.
\cite{Kogut-et-al.90} and we work in the Euclidean theory.
It is well known that:
\be
\displaystyle\lim_{\sup(m_k) {\to} 0} \langle \bar{q}_i q_i \rangle =
-\pi \bar{\rho}(0) ,
\label{EQNE1}
\ee
where $\bar{\rho} (\lambda)$ is simply the gluonic expectation value of 
$\rho (\lambda ;A)$, which is the density of eigenvalues of 
$D[A] \equiv \gamma^\mu D_\mu$ ($D^\mu = \partial^\mu + ig A^\mu$).
In the phase where chiral symmetry is restored we have that:
\be
\displaystyle\lim_{\sup(m_k) {\to} 0} \langle \bar{q}_i q_i \rangle_{T > T_{ch}}
= 0 ,
\label{EQNE2}
\ee
so that $\bar{\rho}(0)=0$: i.e., the ``average'' density of eigenvalues 
of $D[A]$ near $\lambda =0$ is zero in the chirally symmetric phase.
Recall that the partition function becomes ($S_G$ is the gluonic action):
\be
Z = \int [dA]e^{-S_G[A]} \displaystyle\prod_{k=1}^{N_f} \det( D[A] 
+ M_k) ,
\label{EQNE3}
\ee
if we integrate over the $N_f$ quark fields ($L$ of which are taken to be 
``light'' quarks of mass $m_i \ll \Lambda_{QCD}$, $\Lambda_{QCD}$ being the QCD 
mass--scale; the chiral limit is done for this L flavours: $\sup(m_i) \to 0$).
If we are in the chirally symmetric phase, 
then any discrete zero mode due to a non--zero topological charge $Q$ of the 
gauge field $A$ will be well separated from the continuous distribution of
$\bar{\rho} (\lambda)$, since this latter distribution goes to zero at
$\lambda = 0$ by virtue of (\ref{EQNE1}) and (\ref{EQNE2}).
The Atiyah--Singer index theorem relates the number of left--handed 
$n_{-}$ and right--handed $n_{+}$ zero modes of the fermionic operator
$D[A]$ to the topological charge $Q$ of the background gauge field:
\be
n_{+} - n_{-} = Q .
\label{EQNE4}
\ee
So, if we define $n[A]_Q \equiv n_{+} + n_{-} - |Q|$ ($n_{+} + n_{-}$ being 
the total number of discrete zero modes), we find that:
\ba
\lefteqn{
Z = \displaystyle\sum_{Q=0}^\infty \int [dA]_Q e^{-S_G[A]_Q}
\left( \displaystyle\prod_{i=1}^{N_f}M_i \right)^{Q+n[A]_Q} f(M,[A]_Q) = }
\nonumber \\
& & = \displaystyle\sum_{Q=0}^\infty \left( \displaystyle\prod_{i=1}^{N_f}M_i 
\right)^Q \int [dA]_Q e^{-S_G[A]_Q}
\left( \displaystyle\prod_{i=1}^{N_f}M_i \right)^{n[A]_Q} f(M,[A]_Q) =
\nonumber \\
& & = \displaystyle\sum_{Q=0}^\infty \left( \displaystyle\prod_{i=1}^{N_f}M_i 
\right)^Q \alpha_Q(M),
\label{EQNE5}
\ea
where:
\be
\alpha_Q(M) \equiv \int [dA]_Q e^{-S_G[A]_Q}
\left( \displaystyle\prod_{i=1}^{N_f}M_i \right)^{n[A]_Q} f(M,[A]_Q).
\label{EQNE6}
\ee
The $f(M, [A]_Q)$ picks up the contribution of the eigenvalues $\lambda$ 
different from zero: it is of order ${\cal O}(1)$ in the chiral limit
$\sup(m_i) \to 0$. Moreover, it is always possible (at least for $Q=0,
~Q=1$) to find a configuration of the gauge field $A$ in which $n[A]_Q=0$.
All of this implies that $\alpha_Q(M)$ is regular and, in general, non--zero
for $\sup(m_i) \to 0$. If we are in a fixed physical four--volume $V_4$,
then we clearly have:
\ba
\lefteqn{
{1 \over V_4} \langle Q^2 \rangle = { \displaystyle\sum_{Q=0}^\infty Q^2 
\left( \displaystyle\prod_{i=1}^{N_f}{M_i} \right)^Q \alpha_Q(M) \over
V \displaystyle\sum_{Q=0}^\infty {\left( \displaystyle\prod_{i=1}^{N_f}{M_i} 
\right)^Q \alpha_Q(M)}}
\mathop{\longrightarrow}_{\sup(m_k) \to 0} }
\nonumber \\
& & \left( \displaystyle\prod_{i=1}^L{m_i} \right) 
\left( \displaystyle\prod_{k=L+1}^{N_f}{M_k} \right)
{ \alpha_1(0, \ldots ,0,M_{L+1},\ldots ,M_{N_f}) \over V_4
\alpha_0(0,\ldots ,0,M_{L+1},\ldots ,M_{N_f}) } = {\cal O}(\det(M)) ,
\label{EQNE7}
\ea
where $M$ is the usual mass matrix of the $L$ light quarks ($m_i
\equiv M_i$ for $i=1,2,\ldots ,L$).
That is, $\chi = {\cal O}(\prod_{i=1}^L {m_i})$ 
in the chirally symmetric phase. The 
uncertainties of this method lie in the finite--volume approximation and in 
the WKB approximation: probably this is valid here owing to the fact that
$\bar{\rho} (0)=0$. Viceversa, in the region where the chiral symmetry is 
broken we have that $\bar{\rho} (0) \not= 0$, so that any discrete zero 
mode is embedded in a ``sea'' of continuous zero modes. That's why it is not 
immediate to extend the method used before also to the case $T<T_{ch}$.
In Ref. \cite{Hansen-Leutwyler91} the (\ref{EQNE7}) is derived in a different
way, without the WKB approximation, yet always in a finite volume.
With the same technique we can also calculate the chiral condensate
$\langle \bar{q}_i q_i \rangle$. It is clear that:
\ba
\lefteqn{
\langle \bar{q}_i q_i \rangle =
{ \int [dA][dq][d\bar{q}]\bar{q}_i q_i e^{-S_G[A]
-\int d^4x\displaystyle\sum_{k=1}^{N_f}{\bar{q}_k (\gamma^\mu D_\mu +M_k) 
q_k}} \over Z} }
\nonumber \\
& & =-{1 \over V_4 Z}{\partial \over \partial M_i} \int [dA][dq][d\bar{q}]
e^{-S_G[A]
-\int d^4x\displaystyle\sum_{k=1}^{N_f}{\bar{q}_k (\gamma^\mu D_\mu +M_k) 
q_k}}
\nonumber \\
& & =-{1 \over V_4 Z}{\partial Z \over \partial M_i} ,
\label{EQNE8}
\ea
where $V_4$ is always the physical four--volume. Inserting here the (\ref{EQNE5}),
we find that:
\ba
\lefteqn{
\langle \bar{q}_i q_i \rangle = -{1 \over V_4 Z}{\partial \over \partial M_i}
\displaystyle\sum_{Q=0}^\infty { \left( \displaystyle\prod_{k=1}^{N_f}{M_k}
\right) ^Q \alpha_Q (M)} }
\nonumber \\
& & =-{1 \over V_4 Z} \displaystyle\sum_{Q=0}^\infty { \left[ 
\left( \displaystyle\prod_{k=1}^{N_f}{M_k} \right) ^Q
{\partial \alpha_Q (M) \over \partial M_i}
+ QM_i^{Q-1}
\left( \displaystyle\prod_{k \ne i}{M_k} \right) ^Q \alpha_Q (M)
\right] } .
\label{EQNE9}
\ea
In the chiral limit $\sup(m_i) \to 0$, the (\ref{EQNE9}) becomes:
\ba
\lefteqn{
\langle \bar{q}_i q_i \rangle \mathop{\longrightarrow}_{\sup(m_k) \to 0} }
\nonumber \\
& & - \left[ {\partial \alpha_0 \over \partial M_i} +
\left( \displaystyle\prod_{k \ne i}{M_k} \right) 
\alpha_1(0,\ldots ,0,M_{L+1},\ldots ,M_{N_f}) 
\right] \times
\nonumber \\
& & \times { 1 \over V_4 \alpha_0(0,\ldots ,0,M_{L+1},\ldots ,M_{N_f}) } .
\label{EQNE10}
\ea
Considering that, if the operator $D[A]$ has the eigenvalue $\lambda$, it 
also has  the opposite eigenvalue $-\lambda$, one immediately derives
from the (\ref{EQNE6}) that ${\partial \alpha_0 \over \partial M_i} =
{\cal O}(m_i)$. And so we have that:
\be
\langle \bar{q}_i q_i \rangle =
{\cal O}_1 (m_i ) + {\cal O}_2 (\displaystyle\prod_{k \ne i}{m_k}) .
\label{EQNE11}
\ee
The results (\ref{EQNE7}) and (\ref{EQNE11}) are the same that we have found
from our model in the region $T_{ch} < T < T_{U(1)}$ (see Section 14).
Moreover, from the explicit expression of ${\cal O}_2 (\prod_{k \ne i}{m_k})$
given in Eq. (\ref{EQNE10}),
we derive, after a comparison with Eq. (\ref{EQNE7}), the following relation:
\be
{1 \over V_4} \langle Q^2 \rangle =
-m_i {\cal O}_2 (\displaystyle\prod_{k \ne i}{m_k}).
\label{EQNE12}
\ee
(which is, of course, independent of the flavour index ``$i$'').
This is just the Euclidean version of Eq. (\ref{EQN8.12}), derived from our
effective Lagrangian model.

\vfill\eject

{\renewcommand{\Large}{\normalsize}
}

\vfill\eject


\begin{thebibliography}{99}
\bibitem{current-algebra}
S.B. Treiman, in S.B. Treiman, R. Jackiw and D.J. Gross ``{\it Lectures in 
Current Algebra and Its Applications}'' (Princeton University Press,
Princeton, 1972); \\
S.L. Adler and R.F. Dashen, ``{\it Current Algebras}'' (Benjamin, New York,
1968).
\bibitem{quark-model}
M. Gell-Mann and Y. Ne'eman, ``{\it The Eightfold-way}'' (W.A. Benjamin,
New York, 1964); \\
J.J.J Kokkedee, ``{\it The Quark Model}'' (W.A. Benjamin, New York, 1969).
\bibitem{Nambu60}
Y. Nambu, Phys. Rev. Lett. {\bf 4} (1960) 380.
\bibitem{Chou61}
Kuang-chao Chou, Soviet. Phys. JETP {\bf 12} (1961) 492.
\bibitem{Goldstone61}
J. Goldstone, Nuovo Cimento {\bf 19} (1961) 154; \\
S. Coleman, Erice Lectures 1973, in ``{\it Laws of hadronic matter}'',
Academic Press, London and New York (1975); reprinted in: S. Coleman,
``{\it Aspects of symmetry}'', Cambridge Univesity Press (1985).
\bibitem{Blum-et-al.95}
T. Blum, L. Karkk\"ainen, D. Toussaint and S. Gottlieb,
Phys. Rev. D {\bf 51} (1995) 5153.
\bibitem{Weinberg75}
S. Weinberg, Phys. Rev. D {\bf 11} (1975) 3583.
\bibitem{tHooft76}
G. 'tHooft, Phys. Rev. Lett. {\bf 37} (1976) 8; \\
G. 'tHooft, Phys. Rev. D {\bf 14} (1976) 3432.
\bibitem{Witten79a}
E.~Witten, Nucl. Phys. B {\bf 156} (1979) 269.
\bibitem{Veneziano79}
G.~Veneziano, Nucl. Phys. B {\bf 159} (1979) 213.
\bibitem{EM1998}
E. Meggiolaro, Phys. Rev. D {\bf 58} (1998) 085002.
\bibitem{Teper86}
M. Teper, Phys. Lett. B {\bf 171} (1986) 81.
\bibitem{EM1992a}
A. Di Giacomo, E. Meggiolaro, H. Panagopoulos,
Phys. Lett. B {\bf 277} (1992) 491.
\bibitem{EM1995b}
E.-M. Ilgenfritz, E. Meggiolaro and M. M\"uller-Preu{\ss}ker,
Nucl. Phys. B (Proc. Suppl.) {\bf 42} (1995) 496.
\bibitem{Alles-et-al.97}
B. All\'es, M. D'Elia and A. Di Giacomo, Nucl. Phys. B {\bf 494} (1997) 281.
\bibitem{Veneziano80}
G.~Veneziano, Phys. Lett. B {\bf 95} (1980) 90.
\bibitem{Shuryak94}
E. Shuryak, Comments Nucl. Part. Phys. {\bf 21} (1994) 235.
\bibitem{Bernard-et-al.97}
C. Bernard {\it et al.}, Nucl. Phys. B (Proc. Suppl.) {\bf 53} (1997) 442; \\
C. Bernard {\it et al.}, Phys. Rev. Lett. {\bf 78} (1997) 598.
\bibitem{Karsch00}
F. Karsch, Nucl. Phys. B (Proc. Suppl.) {\bf 83--84} (2000) 14.
\bibitem{Vranas00}
P.M. Vranas, Nucl. Phys. B (Proc. Suppl.) {\bf 83--84} (2000) 414.
\bibitem{EM1994a}
E. Meggiolaro, Z. Phys. C {\bf 62} (1994) 669.
\bibitem{EM1994b}
E. Meggiolaro, Z. Phys. C {\bf 62} (1994) 679.
\bibitem{EM1994c}
E. Meggiolaro, Z. Phys. C {\bf 64} (1994) 323.
\bibitem{Nambu-Jona-Lasinio61}
Y. Nambu and G. Jona-Lasinio, Phys. Rev. {\bf 122} (1961) 345; \\
Y. Nambu and G. Jona-Lasinio, Phys. Rev. {\bf 124} (1961) 246.
\bibitem{Carlitz78}
R.D. Carlitz, Phys. Rev. D {\bf 17} (1978) 3225.
\bibitem{Adler-Bell-Jackiw69}
S. Adler, Phys. Rev. {\bf 177} (1969) 2426; \\
J.S. Bell and R. Jackiw, Nuovo Cimento A {\bf 60} (1969) 47.
\bibitem{Belavin-et-al.75}
A.A. Belavin, A.M. Polyakov, A.S. Schwartz and Yu.S. Tyupkin,
Phys. Lett. B {\bf 59} (1975) 85.
\bibitem{tHooft74}
G. 'tHooft, Nucl. Phys. B {\bf 72} (1974) 461.
\bibitem{Veneziano76}
G.~Veneziano, Nucl. Phys. B {\bf 117} (1976) 519.
\bibitem{Isgur76}
N. Isgur, Phys. Rev. D {\bf 13} (1976) 122.
\bibitem{DeRujula-et-al.75}
A. De R\'ujula, H. Georgi and S.L. Glashow, Phys. Rev. D {\bf 12} (1975) 47.
\bibitem{Witten79b}
E. Witten, Nucl. Phys. B {\bf 160} (1979) 57.
\bibitem{Crewther77}
R.J. Crewther, Phys. Lett. B {\bf 70} (1977) 349.
\bibitem{Fujikawa79}
K. Fujikawa, Phys. Rev. Lett. {\bf 42} (1979) 1195.
\bibitem{Briganti-et-al.91}
G. Briganti, A. Di Giacomo and H. Panagopoulos,
Phys. Lett. B {\bf 253} (1991) 427.
\bibitem{EM1992b}
A. Di Giacomo, E. Meggiolaro, H. Panagopoulos,
Phys. Lett. B {\bf 291} (1992) 147.
\bibitem{Pisarski-Wilczek84}
R.D. Pisarski, F. Wilczek, Phys. Rev. D {\bf 29} (1984) 338.
\bibitem{Gross-Pisarski-Yaffe81}
D.J. Gross, R.D. Pisarski and L.G. Yaffe, Rev. Mod. Phys. {\bf 53} (1981) 43.
\bibitem{Kharzeev-et-al.98}
D. Kharzeev, R.D. Pisarski and M.H.G. Tytgat, Phys. Rev. Lett. {\bf 81}
(1998) 512.
\bibitem{EM1995a}
A. Di Giacomo and E. Meggiolaro, Nucl. Phys. B (Proc. Suppl.) {\bf 42}
(1995) 478.
\bibitem{Shifman-Vainshtein-Zakharov79}
M.A. Shifman, A.I. Vainshtein and V.I. Zakharov, 
Nucl. Phys. B {\bf 147} (1979) 385, 448, 519.
\bibitem{Crewther79}
R.J. Crewther, La Rivista del Nuovo Cimento, serie 3, Vol. {\bf 2}, 8 (1979).
\bibitem{DiVecchia-Veneziano80}
P. Di Vecchia and G. Veneziano, Nucl. Phys. B {\bf 171} (1980) 253.
\bibitem{Witten80}
E. Witten, Annals of Physics {\bf 128} (1980) 363.
\bibitem{Rosenzweig-et-al.80}
C. Rosenzweig, J. Schechter and C.G. Trahern,
Phys. Rev. D {\bf 21} (1980) 3388.
\bibitem{Nath-Arnowitt81}
P. Nath and R. Arnowitt, Phys. Rev. D {\bf 23} (1981) 473 .
\bibitem{Leutwyler91}
H.~Leutwyler, ``{\it Chiral Effective Lagrangians}'', BERN preprint, BUTP-91/26 
(1991); lectures given at the ``{\it XXX Internationale universit\"atswochen
f\"ur Kernphysik}'', Schladming (Austria), 1991, and at the ``{\it Advanced
Theoretical Study Institute in Elementary Particle Physics}'', Boulder
(Colorado), 1991.
\bibitem{Teper88}
M.~Teper, Phys. Lett. B {\bf 202} (1988) 553.
\bibitem{Campostrini-SU(3)}
M. Campostrini, A. Di Giacomo, Y. G\"unduc, M.P. Lombardo, H. Panagopoulos 
and R. Tripiccione, Phys. Lett. B {\bf 252} (1990) 436.
\bibitem{Peris94}
S. Peris, Phys. Lett. B {\bf 324} (1994) 442.
\bibitem{Kogut-et-al.90}
J.B. Kogut, D.K. Sinclair, M. Teper, OXFORD preprint, OUTP 90-14P (1990).
\bibitem{Hansen-Leutwyler91}
F.C. Hansen and H. Leutwyler, Nucl. Phys. B {\bf 350} (1991) 201.
\end{thebibliography}
\end{document}